\documentclass[journal,10pt]{IEEEtran}
\usepackage{subfigure}
\PassOptionsToPackage{subfigure}{tocloft}
\usepackage{CJK}
\usepackage{algorithm}
\usepackage{algorithmic}
\usepackage{amsmath}
\usepackage{amssymb}
\usepackage{psfrag}
\usepackage{graphicx}
\usepackage{epstopdf}
\usepackage{multirow}
\usepackage{stfloats}
\usepackage{cite}
\usepackage{color}
\usepackage[normalem]{ulem}
\usepackage{arydshln}
\usepackage{enumerate}
\usepackage{bbm}
\usepackage{bm}
\newcommand{\tabincell}[2]{\begin{tabular}{@{}#1@{}}#2\end{tabular}}

\newtheorem{lemma}{\bf{Lemma}}
\newtheorem{mypro}{\bf{Proposition}}

\title{Weighted Sum-Rate Optimization for Intelligent Reflecting Surface Enhanced Wireless Networks}
\author{
Huayan Guo, \IEEEmembership{Member, IEEE},  Ying-Chang Liang, \IEEEmembership{Fellow, IEEE},\\  Jie Chen, \IEEEmembership{Student Member, IEEE}, and Erik G. Larsson, \IEEEmembership{Fellow, IEEE}
\thanks{This work was supported by the National Natural Science Foundation of China under Grant U1801261, 61631005, and 61571100.  (\textit{Corresponding author: Ying-Chang Liang.})}
\thanks{H. Guo, Y.-C. Liang and J. Chen are with the Center for Intelligent Networking and Communications (CINC), University of Electronic Science and Technology of China (UESTC), Chengdu 611731, China (e-mail: guohuayan@pku.edu.cn; liangyc@ieee.org; chenjie.ay@gmail.com).}
\thanks{E. G. Larsson is with the Department of Electrical Engineering (ISY), Link${\ddot{{\rm o}}}$ping University, SE-581 83 Link${\ddot{{\rm o}}}$ping, Sweden (email: erik.g.larsson@liu.se).}
}

\begin{document}

\maketitle

\begin{abstract}
Intelligent reflecting surface (IRS) is a promising solution to build a programmable wireless environment for future communication systems. In practice, an IRS consists of massive low-cost elements, which can steer the incident signal in fully customizable ways by passive beamforming. In this paper, we consider an IRS-aided multiuser multiple-input single-output (MISO) downlink communication system. In particular, the weighted sum-rate of all users is maximized by joint optimizing the active beamforming at the base-station (BS) and the passive beamforming at the IRS.
In addition, we consider a practical IRS assumption, in which the passive elements can only shift the incident signal to discrete phase levels.
This non-convex problem is firstly decoupled  via Lagrangian dual transform, and then the active and passive beamforming can be optimized alternatingly. The active beamforming at BS is optimized based on the fractional programming method. Then, three efficient algorithms with closed-form expressions are proposed for the passive beamforming at IRS. Simulation results have verified the effectiveness of the proposed algorithms as compared to different benchmark schemes.
\end{abstract}


\section{Introduction}
\emph{Intelligent reflecting surface} (IRS), also known as \emph{large intelligent surface} (LIS), is an artificial  passive radio structure which reflects the incident \emph{radio-frequency} (RF) waves into specified directions with low power consumption \cite{Tan2018SRA,Liu2019metasurface,cuiTJ2017metasurface}.
While IRS resembles  a full-duplex \emph{amplify-and-forward} (AF) relay \cite{Larsson2014FDrelay}, it forwards the RF signals using passive reflection beamforming, thus the power consumption of IRS is much lower than that of the AF relay, and there is nearly no additional thermal noise added during reflecting. Therefore, IRS has recently been considered as the key enabler for smart radio environment, which can greatly enhance the performance of wireless systems \cite{Liaskos2018magzineIRS,Renzo2019position,Hum2014ReflectarrayReview,chenjie2019IRS}.

\begin{figure}
[!t]
\centering
\includegraphics[width=.7\columnwidth]{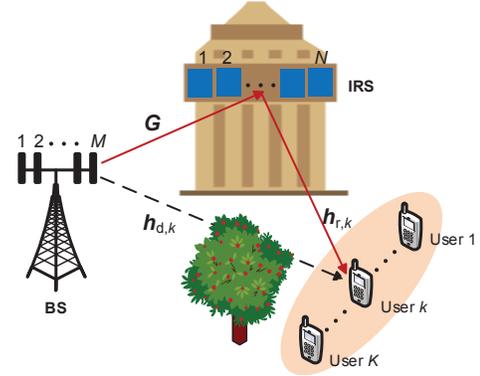}
\caption{An IRS-aided  multiuser MISO communication system.}
\label{IRS_system}
\end{figure}

In this paper, we investigate an IRS-aided \emph{multiple-input single-output} (MISO) multiuser downlink communication system as shown in {\figurename~\ref{IRS_system}}, in which a multi-antenna \emph{base station} (BS) serves multiple single-antenna mobile users.
In the system, the direct links between the BS and the mobile users may suffer from deep fading and shadowing, and the IRS is deployed on a surrounding building's facade to assist the BS in overcoming the unfavorable propagation conditions by providing high-quality virtual links from the BS to the users.
The objective of this paper is to maximize the \emph{weighted sum-rate} (WSR) of the mobile users by jointly optimizing the active beamforming at the BS and the passive beamforming at the IRS.

\subsection{Related Works}
The IRS relays source signals from the BS by passive beamforming, thus the conventional relay beamforming algorithms are not applicable here.
Moreover, the reflection element suffers  a stringent  instantaneous power constraint, which makes the passive beamforming more challenging.
It is worth noting that the joint beamforming problem is much different from the hybrid digital/analog processing  \cite{Ayach2014sparseprecoding,AnLiu2014HybridBeam,YuWei2016HybridBeam} and the constant-envelope precoding \cite{Larsson2013ConstprocodingMU,Larsson2013ConstprocodingFS,Larsson2012SingleuserbeamCEC} in massive \emph{multiple-input multiple-output} (MIMO) systems. Specifically, those designs are restricted to the transceiver sides, while the IRS aims to control and optimize the behavior of the wireless environment.

On the other hand, most existing works on IRS assume that each element is a continuous phase shifter, and then the passive beamforming is equivalent to adjust the phase-shift matrix. In \cite{zhangrui2018GcomIRS} and \cite{zhangruiIRS}, the authors first presented the joint active and passive beamforming problem, while the  transmit power of the BS is minimized based on the \emph{semidefinite relaxation} (SDR) technique.
In \cite{Yuen2018ZF} and \cite{YuenChauIRS}, the authors focused on the maximization of sum-rate and energy efficiency, while employing zero-forcing beamforming at the BS.
In \cite{Dennah2019IRS}, the authors made some practical modifications on the channel model of \cite{YuenChauIRS}, and then the minimum \emph{signal-to-interference-plus-noise ratio} (SINR) of the mobile users is maximized via joint active and passive beamforming. To the best of our knowledge, the joint beamforming for maximizing the WSR of users has not been addressed before.

In practice, the reflection element may only shift the incident signal with discrete \emph{reflection coefficient} (RC) values due to  hardware limitations.
In \cite{Yuen2018EEIRSLowbit}, the authors proposed to quantize the solution of the continuous phase shifter obtained by the function \texttt{fmincon} in MATLAB into the discrete feasible set to maximize the energy efficiency.
However, both the numerical optimization  and the quantization operation are heuristic with unpredictable performance loss.
In \cite{zhangruiIRSDiscrete}, the authors proposed an alternating optimization algorithm to find a local optimal discrete phase-shift solution for the transmit power minimization problem in the single-user MISO system.
However, this method cannot be directly applied to the WSR maximization problem in the multi-user system.

Another key challenge  is the computational complexity.
In practice, the elements on IRS can be massive, thanks to the low cost and low power consumption of the passive components.
Therefore, low-complexity algorithm for passive beamforming is preferred.
In \cite{zhangruiIRS}, the passive beamforming was formulated as a non-convex \emph{quadratically constrained quadratic program} (QCQP), and the SDR technique was employed to solve this problem in polynomial complexity.
However, SDR is not scalable to large-scale IRS as the number of involved variables is quadratic in the number of reflection elements.
In addition, extracting a rank-one component from the optimum solution to the SDR problem is NP-hard in general.
In \cite{Dennah2019IRS}, instead of SDR, the authors  solved the QCQP with low complexity by exploiting the rank-one assumption of the channel between BS and IRS. However, it is not applicable to the general channel model in \cite{zhangruiIRS}.

\subsection{Contributions}
In this paper, we study the joint active and passive beamforming problem to maximize the WSR of the  IRS-aid multiuser downlink MISO system.
This problem is non-convex due to the multiuser interference, and  the optimal solution is unknown.
We try to design an iterative algorithm to find a  suboptimal solution with low computational complexity.
Specifically, we first decouple the active beamforming at BS and the passive beamforming at IRS based on the Lagrangian dual transform proposed in \cite{YuWei2018FP1}.
Then, the active beamforming is solved with closed-form solutions based on the multi-ratio quadratic transform proposed in \cite{YuWei2018FP1}, and the passive beamforming is reformulated as the QCQP which is the same as that in \cite{zhangruiIRS} and \cite{Dennah2019IRS}.

In contrast to \cite{zhangruiIRS} and \cite{Dennah2019IRS}, we attempt to design a unified algorithm for the passive beamforming subproblem, which is applicable to both continuous and discrete phase-shift setups.
To this end, we relax the RC constraint to a ideal convex set, where both the phase and the amplitude of RC can be adjusted.
Then, low-complexity algorithms with closed-form expressions are designed to find the optimal solution of the convex QCQP.
We further show that, these algorithms can be extended to the non-convex phase-shift cases with a small modification.
It is worth noting that, the joint active and passive beamforming solution under the ideal RC assumption not only reveals the ultimate performance limits of the proposed algorithms for the IRS-aided system, but also provides a reasonable initial point for the joint beamforming under non-convex phase-shift assumptions.

The main contributions of this work are summarized as follows:
\begin{itemize}
\item Firstly, this paper is one of the early attempts to study the WSR maximization problem for the IRS-aided multiuser downlink MISO system. An iterative algorithm with closed-form expressions  is proposed to  alternatively optimize the active beamforming at BS and the passive beamforming at IRS.
\item Secondly, we design three low-complexity algorithms for the passive beamforming at IRS. All these algorithms are applicable to the ideal RC assumption, the continuous  phase-shift assumption, as well as the discrete phase-shift assumption.
\item Finally, simulation results have verified that the proposed algorithms achieve significant capacity gains against benchmark schemes. Moreover, the continuous  phase shifter may achieve nearly the same performance as that in ideal cases, and the 2-bit phase shifter may work well with only a small performance degradation.
\end{itemize}

It should be noted that another important application of the passive  radio is ambient backscatter communications \cite{Renzo2019position,Liu2013AmBC,Niyato2018surveyambc,Guo2018IoT,Niyato2017RFCRN} or  symbiotic radio network \cite{GuoAccessSR2019,rzlAccess,Zhang2018JSAC}, which are used to support low-power communications in \emph{Internet of Things} (IoT) applications. In particular, the data of the IoT devices are embedded into the reflected signal from the environment rather than emitting a new radio carrier,  resulting in high spectral and energy efficiency.

The rest of the paper is organized as follows. Section \ref{system model} outlines the system model.
The algorithm framework for joint active and passive beamforming is presented in Section \ref{Sec:WSR}.
In Section \ref{sec:RCopt},  three low-complexity algorithms are proposed for the RC adjustment subproblems.
Simulation results are provided in Section \ref{simulation} to verify the effectiveness of the proposed algorithms, and Section \ref{conclusion} concludes the paper.

The notations used in this paper are listed as follows. ${\mathbb E}[\cdot]$ denotes statistical expectation, $\mathbbm{1} (\cdot)$ is the indicator function, and ${\rm Pj}_{\mathcal F}(\cdot)$ indicates the projection operation onto set ${\mathcal F}$.
${\cal{CN}}(\mu, \sigma^2)$ denotes the \emph{circularly symmetric complex Gaussian} (CSCG) distribution with mean $\mu$ and variance $\sigma^2$.
${\bf{I}}_{M}$ denotes the $M \times M$ identity matrix.
For any general matrix ${\bf G}$, $g_{i,j}$ is the $i$-th row and $j$-th column element.
${\bf G}^{\rm T}$ and ${\bf G}^{\rm H}$ denote the transpose and conjugate transpose of ${\bf G}$, respectively.
For any vector ${\bf w}$ (all vectors in this paper are column vectors), $w_i$ is the $i$-th element, and $\|{\bf w}\|$ denotes the Euclidean norm.
The quantity $\max(x,y)$ and $\min(x,y)$ denote the maximum and minimum between two real numbers $x$ and $y$, respectively.
$|x|$ denotes the absolute value of a complex number $x$, $x^\ast$ denotes its conjugate, and ${\rm{Re}}\{x\}$ and ${\rm{Im}}\{x\}$ denote its real part and imaginary part, respectively.


\section{System Model and Problem Formulation}\label{system model}
\subsection{Channel Model}
{
This paper investigates an IRS-aided multiuser MISO communication system as shown in {\figurename~\ref{IRS_system}}, which consists of one BS equipped with $M$ antennas, one IRS which has $N$ reflection elements, and $K$ single-antenna users.
The baseband equivalent channels from BS to user $k$, from BS to IRS, and from IRS to user $k$ are denoted by ${\bf h}_{{\rm d},k} \in {\mathbb C}^{M\times 1}$, ${\bf G} \in {\mathbb C}^{N\times M}$, and ${\bf h}_{{\rm r},k} \in {\mathbb C}^{N \times 1}$, respectively, with $k=1,\cdots,K$.
For simplicity, we assume that all the channels experience quasi-static flat-fading.
In addition, we assume that the \emph{channel state information} (CSI) of all channels involved is perfectly known at the BS and the IRS, which is the same as \cite{zhangrui2018GcomIRS,zhangruiIRS,Yuen2018ZF,YuenChauIRS,Dennah2019IRS}.

It should be emphasized that the availability of perfect CSI is an idealistic assumption.
Nevertheless, the algorithms proposed under this assumption are still useful as a reference point for studying the theoretical performance gain brought by the IRS, as well as providing training labels for the machine learning based joint beamforming designs, e.g., \cite{Liaskos2019NeuralNetIRS} and \cite{Taha2019IRSChannelE}.
How to obtain CSI at IRS is a difficult task.
Some early-attempts can be found in \cite{Taha2019IRSChannelE} and \cite{Liaskos2019IRSestimation}, in which a channel construct approach is proposed to obtain the full CSI with low training overhead based on compressive sensing tools.

The IRS-aided link (i.e., BS-IRS-user link) is modeled as a concatenation of three components, i.e., the BS-IRS link ${\bf G}$, IRS phase-shift matrix (i.e., passive beamforming), and IRS-user link ${\bf h}_{{\rm d},k}$.
Denote by $\theta_n \in {\mathcal F}$ the RC of the $n$-th reflection element, where ${\mathcal F}$ is the feasible set of RC.
The reflection operation on the IRS element resembles multiplying the incident signal with $\theta_n$, and then forwarding this composite signal as if from a point source, which is the main difference from the active reflection  surface \cite{Edfors2018IRSMIMOactive,Saad2019activeLISUplink,Saad2018activeLISrate}.
It is known that, the power of signals decreases drastically during reflection.\footnote{The power loss of reflection operation is generally larger than $10$ dB due to the ``double-fading'' effect \cite{Griffin2009LinkBudget} which will be presented in the link budget in Section \ref{simulation_link}.}
Thus the phase-shift matrix of IRS is approximately denoted by a  diagonal matrix ${\bf \Theta}=\sqrt{\eta} {\rm diag}(\theta_1, \cdots, \theta_n, \cdots, \theta_N)$ (where $\eta\leq 1$ indicates the reflection efficiency), for the power of signals reflected two or more times is much smaller than that of the signal reflected only one time.
Besides, we consider following three assumptions for the feasible set of RC in this paper:
}
\begin{itemize}
\item \emph{Ideal RC}: Under this assumption, we only restrict that the RC is peak-power constrained:
\begin{equation}
{\mathcal F}_1 = \left\{\theta_n \left| |\theta_n|^2 \leq 1 \right. \right\}.
\end{equation}
It is shown in \cite{LiuFu2019Metasurface} that, the amplitude and phase of $\theta_n$ can be controlled independently via controlling over the resistance and capacitance of the integrated circuits in the IRS element, respectively.
Under this assumption, the theoretical performance upper bound of passive beamforming can be obtained afterwards.
\item \emph{Continuous Phase Shifter}: In \cite{zhangrui2018GcomIRS,zhangruiIRS,Yuen2018ZF,YuenChauIRS,Dennah2019IRS}, it is assumed that the strength of the reflection signal from each reflection element is maximized, thus $|\theta_n|^2 = 1$. Then, the reflection element only adjusts the phase of the incident signal, and we have $\theta_n= e^{j\varphi_n}$. Since $\theta_n$ can be adjusted to any desired phase, we have:
\begin{equation}
{\mathcal F}_2 = \left\{\theta_n \left| \theta_n= e^{j\varphi_n}, \varphi_n \in [0, 2 \pi)  \right. \right\}.
\end{equation}
\item \emph{Discrete Phase Shifter}: In practice, the  reflection element only has finite reflection levels. Same as \cite{Yuen2018EEIRSLowbit} and \cite{zhangruiIRSDiscrete}, we assume that $\theta_n$ only takes $\tau$ discrete values  which are equally spaced on the circle $\theta_n= e^{j\varphi_n}$, i.e.,
\begin{equation}
{\mathcal F}_3 = \left\{\theta_n \left| \theta_n= e^{j\varphi_n}, \varphi_n  \in \{0, \frac{2 \pi}{\tau}, \cdots, \frac{2 \pi (\tau-1)}{\tau} \}  \right. \right\}.
\end{equation}
\end{itemize}

\subsection{Received Signal at User $k$}
Denote the transmit data symbol to user $k$ by $s_k$. It is assumed that $s_k$ ($k=1, \cdots, K$) are independent random variables with zero mean and unit variance.
Then, the transmitted signal at the BS can be expressed as
\begin{equation}
{\bf x}=\sum_{k=1}^K {\bf w}_k s_k,
\end{equation}
where ${\bf w}_k \in {\mathbb C}^{M\times 1}$ is the corresponding transmit beamforming vector.

The signal received at user $k$ is expressed as
\begin{equation}
\begin{aligned}[b]
{y}_k&= \underbrace{ {\bf h}_{{\rm d},k}^{\rm H} {\bf x} }_{\rm {Direct~link}}
+\underbrace{ {\bf h}_{{\rm r},k}^{\rm H} {\bf \Theta}^{\rm H} {\bf G}  {\bf x} }_{{\rm IRS}-{\rm aided}~{\rm link}}
+u_k\\
&=\left({\bf h}_{{\rm d},k}^{\rm H}+{\bf h}_{{\rm r},k}^{\rm H} {\bf \Theta}^{\rm H} {\bf G}  \right) \sum_{k=1}^K {\bf w}_k s_k +u_k
,
\end{aligned}
\end{equation}
where $u_k \sim {\cal{CN}}(0,\sigma_0^2) $ denotes the \emph{additive white Gaussian noise} (AWGN) at the $k$-th user receiver.

\subsection{Problem Formulation}
The $k$-th user  treats all the signals from other users (i.e., $s_1,\cdots,s_{k-1},s_{k+1},\cdots,s_K$) as interference.
Hence, the decoding SINR of $s_k$ at user $k$ is
\begin{equation}\label{equ:downlink_SINR}
{\gamma}_k= \frac{\left|({\bf h}_{{\rm d},k}^{\rm H}+{\bf h}_{{\rm r},k}^{\rm H} {\bf \Theta}^{\rm H} {\bf G}){\bf w}_k  \right|^2 }
{\sum_{i=1, i \neq k}^K \left|({\bf h}_{{\rm d},k}^{\rm H}+{\bf h}_{{\rm r},k}^{\rm H} {\bf \Theta}^{\rm H} {\bf G}){\bf w}_i  \right|^2+\sigma_0^2}
.
\end{equation}
The transmit power constraint of BS is
\begin{equation}\label{equ:power_constrant_BS}
\sum_{k=1}^K \|{\bf w}_k\|^2 \leq P_{\rm T}
.
\end{equation}

Let ${\bf W}=[{\bf w}_1,{\bf w}_2, \cdots, {\bf w}_K] \in {\mathbb C}^{M\times K}$.
In this paper, our objective is to maximize the WSR of all the users by jointly designing the transmit beamforming matrix ${\bf W}$ at the BS and the RC matrix  ${\bf \Theta}$ at IRS, subject to the transmit power constraint in \eqref{equ:power_constrant_BS}.
The WSR maximization problem is thus formulated as
\begin{subequations}
\begin{align}
({\rm P}1)\quad \max_{\bf { W, \Theta}} \quad &  f_{1}({\bf { W, \Theta}})=\sum_{k=1}^K \omega_k \log_2(1+{\gamma}_k) \notag \\
{\bf s.t.} \quad
& \theta_n \in {\mathcal F}, \quad \forall n=1,\cdots,N, \label{equ:P1c1}\\
& \sum_{k=1}^K \|{\bf w}_k\|^2 \leq P_{\rm T}, \label{equ:P1c2}
\end{align}
\end{subequations}
where ${\mathcal F} \in \{{\mathcal F}_1, {\mathcal F}_2, {\mathcal F}_3\}$, and the weight $\omega_k$ is used to represent the priority of user $k$.

Despite the conciseness of $({\rm P}1)$, it is generally much more difficult than the power minimization problem in \cite{zhangruiIRS}
due to the non-convex objective function $f_{1}({\bf { W, \Theta}})$ and  the non-convex constraint sets ${\mathcal F}_2$ and ${\mathcal F}_3$.\footnote{Mathematically, the problem of minimizing the transmit power given individual rate requirements of users is equivalent to the problem of maximizing the minimum SINR of users given transmit power constraint.}
In this paper, we try to find a suboptimal solution for $({\rm P}1)$ with low computational complexity.
To be specific, we need to address two technical challenges:
\begin{itemize}
\item First, we need to decouple the optimization variables in $f_{1}({\bf { W, \Theta}})$ to make $({\rm P}1)$ non-convex and intractable.
\item Second,  the complexity for RC adjustment algorithm should be scalable for cases with  large $N$.
\end{itemize}

\section{WSR Maximization for Downlink Transmission}\label{Sec:WSR}
In this section, we address the first challenge to decouple the optimization of the transmit beamforming $\bf W$  and the RC  matrix $\bf \Theta$ into several tractable subproblems.

\subsection{Lagrangian Dual Transform}
To tackle the logarithm in the objective function of $({\rm P}1)$, we apply the the Lagrangian dual transform proposed in \cite{YuWei2018FP1}.
Then, (P1) can be equivalently written as
\begin{align*}
({\rm P}1')\quad \max_{{\bf W, \Theta}, {\bm \alpha}} \quad &   f_{1a}({\bf  W, \Theta}, {\bm \alpha}) \notag \\
{\bf s.t.} \quad
& \eqref{equ:P1c1}, \eqref{equ:P1c2}, \notag
\end{align*}
where $\bm \alpha$ refers to $[\alpha_1,\cdots, \alpha_k, \cdots, \alpha_K]^{\rm T}$, and $\alpha_k$ is an auxiliary variable for the decoding SINR $\gamma_k$; and the new objective function is defined by
\begin{equation}\label{equ:obj_P1n}
\begin{aligned}[b]
f_{1a}({\bf  W, \Theta}, {\bm \alpha})&=\sum_{k=1}^K \omega_k \log_2(1+\alpha_k)-\sum_{k=1}^K \omega_k \alpha_k \\
&\qquad + \sum_{k=1}^K \frac{\omega_k (1+\alpha_k) {\gamma}_k}{1+{\gamma}_k}.
\end{aligned}
\end{equation}

In $({\rm P}1')$, when $\bf W$ and $\bf \Theta$ hold fixed, the optimal $\alpha_k$ is
\begin{equation}\label{equ:uplink_gamma}
\alpha_k^\circ= \gamma_k
.
\end{equation}
Then, for a fixed $\bm \alpha$, optimizing $\bf W$ and $\bf \Theta$ is reduced to
\begin{align*}
({\rm P}1'')\quad \max_{{\bf  W, \Theta}} \quad &   \sum_{k=1}^K \frac{{\tilde{\alpha}}_k {\gamma}_k}{1+{\gamma}_k} \notag \\
{\bf s.t.} \quad
& \eqref{equ:P1c1}, \eqref{equ:P1c2}, \notag
\end{align*}
where ${\tilde{\alpha}}_k=\omega_k (1+\alpha_k)$.

\begin{figure}
[!t]
\centering
\includegraphics[width=.9\columnwidth]{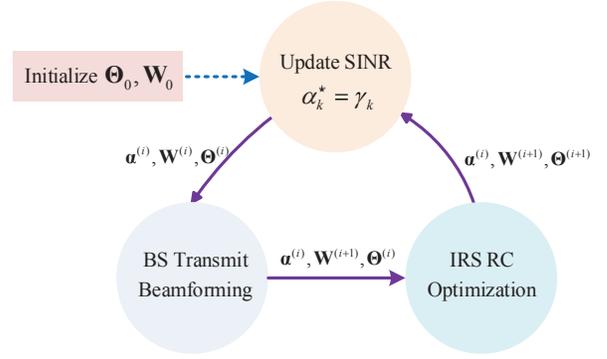}
\caption{Alternating optimization for $({\rm P}1')$.}
\label{AO}
\end{figure}

$({\rm P}1'')$ is the sum of multiple-ratio FP problems, and the non-convexity introduced by the ratio operation can be solved via the recently proposed  fractional programming technique \cite{YuWei2018FP1}.
In the next two subsections, we will investigate how to solve $\bf W$ by fixing $\bf \Theta$ and to solve $\bf \Theta$ by fixing $\bf W$, respectively.
Then, the original problem $({\rm P}1')$ can be solved in an iterative manner by applying the alternating optimization as illustrated in {\figurename~\ref{AO}}.
In particular, in each iteration, we first update the nominal SINR ${\bm \alpha}$, and then better solutions for $\bf W$ and $\bf \Theta$ are updated, respectively. The process is repeated until no further improvement is obtained.

\subsection{Transmit Beamforming}
In this subsection, we investigate how to find a better beamforming matrix $\bf W$ given fixed $\bf \Theta$ for $({\rm P}1'')$.
Denote the combined channel for user $k$ by
\begin{equation}
{\bf h}_k={\bf h}_{{\rm d},k}+{\bf G}^{\rm H} {\bf \Theta} {\bf h}_{{\rm r},k} .
\end{equation}
Then, the SINR ${\gamma}_k$ in \eqref{equ:downlink_SINR} becomes
\begin{equation}\label{equ:downlink_SINR2}
{\gamma}_k= \frac{\left|{\bf h}_k^{\rm H} {\bf w}_k  \right|^2 }
{\sum_{i=1, i \neq k}^K \left|{\bf h}_k^{\rm H} {\bf w}_i  \right|^2+\sigma_0^2}
.
\end{equation}
Using ${\gamma}_k$ in \eqref{equ:downlink_SINR2}, the objective function of $({\rm P}1'')$ is written as a function of $\bf  W$:
\begin{equation}\label{equ:f2}
\begin{aligned}[b]
f_2({\bf  W})&=
\sum_{k=1}^K \frac{ {\tilde{\alpha}}_k {\gamma}_k}{1+{\gamma}_k}\\
&=\sum_{k=1}^K  \frac{ {\tilde{\alpha}}_k \left|{\bf h}_k^{\rm H} {\bf w}_k  \right|^2 }
{\sum_{i=1}^K \left|{\bf h}_k^{\rm H} {\bf w}_i  \right|^2+\sigma_0^2}
.
\end{aligned}
\end{equation}
Thus, given $\bm \alpha$ and $\bf \Theta$, optimizing $\bf W$ becomes
\begin{align*}
({\rm P}2)\quad \max_{{\bf  W}} \quad &   f_2({\bf  W}) \notag \\
{\bf s.t.} \quad
&  \sum_{k=1}^K \|{\bf w}_k\|^2 \leq P_{\rm T}.
\end{align*}

It is known that, $({\rm P}2)$ is the multiple-ratio fractional programming problem.
Using quadratic transform proposed in \cite{YuWei2018FP1}, $f_2({\bf  W})$ is reformulated as
\begin{equation}\label{equ:quadratic1}
\begin{aligned}[b]
f_{2a}({\bf W},{\bm \beta})&= \sum_{k=1}^K
2 \sqrt{{\tilde{\alpha}}_k}  {\rm Re} \left\{ \beta_k^{\ast} {\bf h}_k^{\rm H} {\bf w}_k \right\}\\
&\quad-\sum_{k=1}^K \left|\beta_k \right|^2 \left(\sum_{i=1}^K \left|{\bf h}_k^{\rm H} {\bf w}_i  \right|^2+\sigma_0^2\right)
.
\end{aligned}
\end{equation}
where ${\bm \beta}=[\beta_1, \cdots, \beta_K]^{\rm T}$, and $\beta_k\in {\mathbb C}$ is the auxiliary variable.
Then, based on \cite{YuWei2018FP1}, solving problem $({\rm P}2)$ over ${\bf W}$ is equivalent to solving the following problem over ${\bf W}$ and ${\bm \beta}$:
\begin{align*}
({\rm {P}}2a) \: \max_{{\bf W}, {\bm \beta}} \quad &  f_{2a}({\bf W},{\bm \beta})
 \notag \\
{\bf s.t.} \quad
&  \sum_{k=1}^K \|{\bf w}_k\|^2 \leq P_{\rm T}.
\end{align*}
$({\rm {P}}2a)$ is a biconvex optimization problem, and a common practice for solving it (which does not guarantee global optimality of the solution) is alternatively updating ${\bf W}$ and ${\bm \beta}$ by fixing one of them and solving the corresponding convex optimization problem \cite{Jochen2007biconvex}.

\begin{lemma}
The optimal $\beta_k$ for a given ${ {\bm W}}$ is
\begin{equation}\label{equ:opt_beta}
\beta_k^\circ = \frac{ \sqrt{{\tilde{\alpha}}_k } {\bf h}_k^{\rm H} {\bf w}_k }
{\sum_{i=1}^K \left|{\bf h}_k^{\rm H} {\bf w}_i  \right|^2+\sigma_0^2}
.
\end{equation}
Then, fixing ${\bm \beta}$, the optimal ${\bf w}_k$ is
\begin{equation}\label{equ:opt_w}
{\bf w}_k^\circ=\sqrt{{\tilde{\alpha}}_k } \beta_k
\left({\lambda_0 {\bf I}_M+\sum_{i=1}^K |\beta_i|^2 {\bf h}_i {\bf h}_i^{\rm H} }\right)^{-1}
{   {\bf h}_k }
,
\end{equation}
where $\lambda_0$ is the dual variable introduced for the power constraint, which is optimally determined by
\begin{equation}
\lambda_0^\circ=\min \left\{
   \lambda_0 \geq 0 : \sum_{k=1}^K \|{\bf w}_k\|^2 \leq P_{\rm T}
\right\}
.
\end{equation}
\end{lemma}

\begin{IEEEproof}
$\beta_k^\circ$ in \eqref{equ:opt_beta} and ${\bf w}_k^\circ$ in \eqref{equ:opt_w} can be obtained by setting $\partial f_{2a}/\partial \beta_k$ and $\partial f_{2a}/\partial {\bf w}_k$ to zero, respectively.
\end{IEEEproof}

\subsection{Optimizing Reflection Response Matrix $\bf \Theta$}
Finally, we optimize $\bf \Theta$ in $({\rm P}1'')$  given  fixed $\bm \alpha$ and $\bf W$.
Using $\gamma_k$ defined in \eqref{equ:downlink_SINR}, the objective function of $({\rm P}1'')$ is expressed as a function of $\bf \Theta$:
\begin{equation}\label{equ:f3}
\begin{aligned}[b]
f_{3u}({\bf \Theta})
&=\sum_{k=1}^K \frac{ {\tilde{\alpha}}_k {\gamma}_k}{1+{\gamma}_k}\\
&=\sum_{k=1}^K \frac{{\tilde{\alpha}}_k {|({\bf h}_{{\rm d},k}^{\rm H}+{\bf h}_{{\rm r},k}^{\rm H} {\bf \Theta}^{\rm H} {\bf G}){\bf w}_k |^2 }}
{\sum_{i=1}^K |({\bf h}_{{\rm d},k}^{\rm H}+{\bf h}_{{\rm r},k}^{\rm H} {\bf \Theta}^{\rm H} {\bf G}){\bf w}_i |^2 +\sigma_0^2}
.
\end{aligned}
\end{equation}

Define ${\bf a}_{i,k}=\sqrt{\eta} {\rm diag}({\bf h}_{{\rm r},k}^{\rm H}){\bf G}{\bf w}_i $, $b_{i,k}= {\bf h}_{{\rm d},k}^{\rm H} {\bf w}_i$, and ${\bm \theta}= [\theta_1, \cdots, \theta_N]^{\rm T}$. Then, $| ({\bf h}_{{\rm d},k}^{\rm H}+ {\bf h}_{{\rm r},k}^{\rm H} {\bf \Theta}^{\rm H}{\bf G} ) {\bf w}_i|^2$ in \eqref{equ:f3} becomes
\begin{equation}\label{equ:uplink_P1c_getv}
\begin{aligned}[b]
| ({\bf h}_{{\rm d},k}^{\rm H}+ {\bf h}_{{\rm r},k}^{\rm H} {\bf \Theta}^{\rm H}{\bf G} ) {\bf w}_i|^2
&=|b_{i,k}+\sqrt{\eta}  {\bm \theta}^{\rm H} {\rm diag}({\bf h}_{{\rm r},k}^{\rm H}){\bf G}{\bf w}_i|^2\\
&=|b_{i,k}+{\bm \theta}^{\rm H}  {\bf a}_{i,k}|^2
,
\end{aligned}
\end{equation}
for all $i$ and $k$.
Using \eqref{equ:uplink_P1c_getv}, $f_{3u}({\bf \Theta})$ in \eqref{equ:f3} is equivalently transformed to a new function of ${\bm \theta}$:
\begin{equation}
f_{3}({\bm \theta})=\sum_{k=1}^K  \frac{ {\tilde{\alpha}}_k |b_{k,k}+{\bm \theta}^{\rm H}  {\bf a}_{k,k}|^2 }
{\sum_{i=1}^K |b_{i,k}+{\bm \theta}^{\rm H}  {\bf a}_{i,k}|^2 +\sigma_0^2}
.
\end{equation}
Finally, optimizing $\bf \Theta$ is translated  to optimizing ${\bm \theta}$, which is represented as follows:
\begin{align*}
({\rm P}3)\: \max_{{\bm \theta}} \quad & f_{3}({\bm \theta})
\notag \\
{\bf s.t.} \quad & \theta_n \in {\mathcal F}, \quad \forall n=1,\cdots,N.
\end{align*}

$({\rm P}3)$ is also a multiple-ratio fractional programming problem, and can be translated to the following problem based on the quadratic transform proposed in \cite{YuWei2018FP1}:
\begin{align*}
({\rm P}3a)\: \max_{{ {\bm \theta}}, \bm \varepsilon} \quad &
f_{3a}({ {\bm \theta}}, \bm \varepsilon)
\notag \\
{\bf s.t.} \quad & \theta_n \in {\mathcal F}_{\rm D}, \quad \forall n=1,\cdots,N,
\end{align*}
where the new objective function is
\begin{equation}\label{equ:quadratic2}
\begin{aligned}[b]
f_{3a}({ {\bm \theta}}, \bm \varepsilon)&=
\sum_{k=1}^K    2 \sqrt{{\tilde{\alpha}}_k }
{\rm Re} \left\{  \varepsilon_k^{\ast} { {\bm \theta}}^{\rm H}  { {\bf a}}_{k,k}+ \varepsilon_k^{\ast} b_{k,k}\right\} \\
&\quad-\sum_{k=1}^K \left|\varepsilon_k\right|^2
\left(\sum_{i=1}^K |b_{i,k}+{\bm \theta}^{\rm H}  {\bf a}_{i,k}|^2 +\sigma_0^2\right)
,
\end{aligned}
\end{equation}
and ${\bm \varepsilon}$ refers to the auxiliary variable vector $[\varepsilon_1, \cdots, \varepsilon_K]^{\rm T}$.

Similarly, we optimize ${ {\bm \theta}}$ and $\bm \varepsilon$ alternatively.
The optimal $\varepsilon_k$ for a given ${ {\bm \theta}}$ can be obtained by setting $\partial f_{3a}/\partial \varepsilon_k$ to zero, i.e.,
\begin{equation}\label{equ:opt_epsilon}
\varepsilon_k^\circ = \frac{\sqrt{{\tilde{\alpha}}_k } \left(b_{k,k}+{\bm \theta}^{\rm H}  {\bf a}_{k,k}\right) }
{{\sum_{i=1}^K |b_{i,k}+{\bm \theta}^{\rm H}  {\bf a}_{i,k}|^2 +\sigma_0^2}}
.
\end{equation}
Then the remaining problem is optimizing ${ {\bm \theta}}$ for a given $\bm \varepsilon$.
It is known that, $|b_{i,k}+{\bm \theta}^{\rm H}  {\bf a}_{i,k}|^2$ in \eqref{equ:quadratic2} can be further written as
\begin{equation}\label{equ:A1_f4_disc}
\begin{aligned}[b]
|b_{i,k}+{\bm \theta}^{\rm H}  {\bf a}_{i,k}|^2&
=\left(b_{i,k}+{\bm \theta}^{\rm H}  {\bf a}_{i,k}\right)\left(b_{i,k}^{\ast}+ {\bf a}_{i,k}^{\rm H} {\bm \theta} \right)\\
&={\bm \theta}^{\rm H}  {\bf a}_{i,k}{\bf a}_{i,k}^{\rm H} {\bm \theta}
+2 {\rm Re} \left\{ b_{i,k}^{\ast}{\bm \theta}^{\rm H}  {\bf a}_{i,k} \right\}+|b_{i,k}|^2
.
\end{aligned}
\end{equation}
Substituting \eqref{equ:opt_epsilon} and \eqref{equ:A1_f4_disc} into \eqref{equ:quadratic2}, the  optimization problem for ${ {\bm \theta}}$ is represented as follows
\begin{equation}\label{equ:P4c1}
\begin{aligned}[b]
({\rm P}4)\: \max_{{ {\bm \theta}}} \quad &
f_{4}({ {\bm \theta}})
\\
{\bf s.t.} \quad & \theta_n \in {\mathcal F}_{\rm D}, \quad \forall n=1,\cdots,N,
\end{aligned}
\end{equation}
where the objective function is
\begin{equation}\label{equ:f4bcro}
\begin{aligned}[b]
f_{4}({ {\bm \theta}})
&=f_{3a}({ {\bm \theta}}, {\bm \varepsilon}^\circ)\\
&=-{\bm \theta}^{\rm H}  {\bm U} {\bm \theta}
+ 2 {\rm Re} \left\{ {\bm \theta}^{\rm H} {\bm \nu}
\right\}+C
,
\end{aligned}
\end{equation}
and
\begin{align}
{\bm U}&=
\sum_{k=1}^K \left|\varepsilon_k\right|^2
\sum_{i=1}^K  {\bf a}_{i,k}{\bf a}_{i,k}^{\rm H}, \label{equ:f4U}
 \\
{\bm \nu}&= \sum_{k=1}^K  \left(  \sqrt{{\tilde{\alpha}}_k } \varepsilon_k^{\ast} { {\bf a}}_{k,k}
    -  \left|\varepsilon_k\right|^2 \sum_{i=1}^K   b_{i,k}^{\ast}  {\bf a}_{i,k}
\right), \label{equ:f4nu}\\
C&=\sum_{k=1}^K  \left(  2 \sqrt{{\tilde{\alpha}}_k }
{\rm Re} \left\{ \varepsilon_k^{\ast} b_{k,k}\right\}
- \left|\varepsilon_k\right|^2 (\sigma_0^2+ \sum_{i=1}^K |b_{i,k}|^2)
\right)
.
\end{align}

Since ${\bf a}_{i,k}{\bf a}_{i,k}^{\rm H}$ for all $i$ and $k$ are positive-definite matrices, ${\bm U}$ is a positive-definite matrix, and $f_{4}({ {\bm \theta}})$ is a quadratic concave function of $\bm \theta$.
Therefore, the passive beamforming subproblem $({\rm P}4)$ is a QCQP which is the same as that  in \cite{zhangruiIRS} and \cite{Dennah2019IRS},
and the non-convexity of $({\rm P}4)$ is only introduced by the constraint in \eqref{equ:P4c1}.
We will investigate the algorithms to solve  $({\rm P}4)$ in the next section.

\subsection{Algorithm Development}
We summarize the proposed alternating optimization method in {\emph{Algorithm} \ref{alg:P1}}.
To be specific, the algorithm starts with certain feasible values of ${ {\bm W}}^{(0)}$ and ${\bm \Theta}^{(0)}$.
Next, given a fixed solution $\{ {\bm W}^{(i)}, {\bm \Theta}^{(i)} \}$ in the $i$-th iteration, we first update the the nominal SINR ${\bm \alpha}^{(i+1)}$, and then the transmit beamforming  ${\bm W}^{(i+1)}$ and RC values of IRS ${\bm \Theta}^{(i+1)}$ are updated based on the  fractional programming techniques, respectively, for the $(i+1)$-th iteration.
The convergence of the whole algorithm is discussed in the following proposition.

\begin{mypro}\label{pro_converge}
{\emph{Algorithm} \ref{alg:P1}} is guaranteed to converge, if the RC vector ${\bm \theta}$ obtained by solving $({\rm P}4)$ in the $i$-th iteration satisfies:
\begin{equation}\label{equ:con_pro1}
f_{4}({ {\bm \theta}^{(i)} }) \geq f_{4}( {\bm \theta}^{(i-1)}).
\end{equation}
\end{mypro}

\begin{IEEEproof}
It can be verified that, when \eqref{equ:con_pro1} is satisfied, the objective function is monotonically nondecreasing after each iteration. Therefore, {\emph{Algorithm} \ref{alg:P1}} is guaranteed to converge.
\end{IEEEproof}

\begin{algorithm}[!ht]
\caption{The  alternating optimization for solving $({\rm P}1')$.}
\label{alg:P1}
\begin{algorithmic}[1]
\STATE {\bf Step 0}: {Initialize ${ {\bm W}}^{(0)}$ and ${\bm \Theta}^{(0)}$ to feasible values.\\
{\bf Repeat}}
\STATE {\bf Step 1}: Update the nominal SINR ${\bm \alpha}^{(i)}$ by \eqref{equ:uplink_gamma};
\STATE {\bf Step 2.1}: Update ${\bm \beta}^{(i)}$ by \eqref{equ:opt_beta};
\STATE {\bf Step 2.2}: Update transmit beamforming  ${\bm W}^{(i)}$ by \eqref{equ:opt_w};
\STATE {\bf Step 3.1}: Update ${\bm \varepsilon}^{(i)}$ by \eqref{equ:opt_epsilon};
\STATE {\bf Step 3.2}: Update RC values in ${\bm \Theta}^{(i)}$ by solving $({\rm P}4)$;
\\
{\bf Until} The value of function $f_{1a}$ in \eqref{equ:obj_P1n} converges.
\end{algorithmic}
\end{algorithm}

\section{Reflection Coefficients Adjustment for $({\rm P}4)$}\label{sec:RCopt}
In Section \ref{Sec:WSR}, the WSR maximization problem is decoupled, and an iterative algorithm for joint active and passive beamforming is proposed.
We have derived closed-form solutions for every step in {\emph{Algorithm} \ref{alg:P1}} except {\emph {Step}} 3.2, i.e., optimizing the RC by solving $({\rm P}4)$.
Thus we address $({\rm P}4)$ in this section.

After dropping irrelevant constant terms,  $({\rm P}4)$ is equivalently translated to
\begin{equation}\label{equ:P4bc1}
\begin{aligned}[b]
({\rm P}4a)\: \max_{{ {\bm \theta}}} \quad &
f_{4a}({ {\bm \theta}})
\\
{\bf s.t.} \quad & |\theta_n|^2 \in {\mathcal F}, \quad \forall n=1,\cdots,N,
\end{aligned}
\end{equation}
where
\begin{equation}\label{equ:f4bcr}
\begin{aligned}[b]
&f_{4a}({ {\bm \theta}})
=-{\bm \theta}^{\rm H}  {\bm U} {\bm \theta}
+ 2 {\rm Re} \left\{ {\bm \theta}^{\rm H} {\bm \nu}
\right\}
.
\end{aligned}
\end{equation}
Note that, $f_{4a}({ {\bm \theta}})$ is a  concave quadratic function.
When ${\theta_n} \in {\mathcal F}_1$, the constraints in \eqref{equ:P4bc1} is also convex. Hence $({\rm P}4a)$ is convex in this case.
However, when  ${\theta_n} \in {\mathcal F}_2$ (or ${\theta_n} \in {\mathcal F}_3$), $({\rm P}4a)$ is non-convex, and to find the optimal ${\bm \theta}$ is a challenging task.

In \cite{zhangruiIRS}, the author applied SDR to solve this problem. Later they used Gaussian randomization to construct a rank-one solution.
In general, the computation complexity of SDR is in the order of ${\mathcal O}(N^6)$ \cite{LectConvexOpt2001BEN}, which is not scalable for cases with large $N$.
Therefore, in this section, we aim to design low-complexity and scalable algorithms to solve $({\rm P}4)$.
In particular, we first propose algorithms with closed-form solutions for the convex case (i.e., ${\theta_n} \in {\mathcal F}_1$).
Then, the algorithms proposed for the convex problem are extended to tackle the non-convex cases.

\subsection{Nearest Point Projection}\label{theta_LDD}
In \cite{Yuen2018EEIRSLowbit}, the authors suggested solving the RC adjustment problem for ${\mathcal F}_2$ first, and then projecting the solution to ${\mathcal F}_3$.
The performance of the projection solution is highly related to that of the solution for the original problem.
However, since ${\mathcal F}_2$ is non-convex, the optimal solution under this feasible set is difficult to obtain.
To overcome this drawback, we make a small modification on that two-step method:
\begin{itemize}
\item Firstly, we solve the convex problem by assuming ${\theta_n} \in {\mathcal F}_1$, and derive the optimal  ${\bm \theta}$.
\item Secondly, ${\bm \theta}$ is projected to the nearest feasible point in the non-convex set ${\mathcal F}_2$ or ${\mathcal F}_3$ to obtain a suboptimal solution for the non-convex problem.
\end{itemize}
We name this method as the \emph{nearest point projection} (NPP) method.
The projection solution in our method may achieve better performance, since at the first step, we obtain the optimal solution instead of a suboptimal one carried out by numerical optimization.


\subsubsection{Convex Optimization Step}
When $\theta_n \in {\mathcal F}_1$, the constraint in \eqref{equ:P4c1} becomes
\begin{equation}
|\theta_n|^2 \leq 1, \quad \forall n=1,\cdots,N.
\end{equation}
However,  $f(\theta)=|\theta|^2$ is not a complex analytic function. Thus, we rewrite the above constraint as
\begin{equation}\label{equ:f4nconst1}
{\bm \theta}^{\rm H} {\bf e}_n {\bf e}_n^{\rm H} {\bm \theta}\leq 1, \quad \forall n=1,\cdots,N,
\end{equation}
where ${\bf e}_n \in {\mathbb R}^{N \times 1}$ is an elementary vector with a one at the $n$-th position.
Then, $({\rm P}4a)$ (with ${\mathcal F}={\mathcal F}_1$) is represented as
\begin{align*}
({\rm P}4a)\: \max_{{ {\bm \theta}}} \quad &
f_{4a}({ {\bm \theta}})
\notag \\
{\bf s.t.} \quad & {\bm \theta}^{\rm H} {\bf e}_n {\bf e}_n^{\rm H} {\bm \theta}\leq 1, \quad \forall n=1,\cdots,N.
\end{align*}
The above problem is convex, and it can be equivalently transformed to the dual problem via \emph{Lagrange dual decomposition} (LDD):
\begin{align*}
({\rm P}4b)\: \min_{ {\bm \lambda} } \quad &
{\mathcal L} ({\bm \lambda})=\max_{{ {\bm \theta}}} \left\{
 {\mathcal G}({\bm \theta},{\bm \lambda})
\right\}
\notag \\
{\bf s.t.} \quad & \lambda_n \geq 0, \quad \forall n=1,\cdots,N,
\end{align*}
where ${\bm \lambda}=[\lambda_1,\cdots,\lambda_N]$, $\lambda_n$ is the dual variable for the constraint ${\bm \theta}^{\rm H} {\bf e}_n {\bf e}_n^{\rm H} {\bm \theta}\leq 1$, and ${\mathcal L} ({\bm \theta},{\bm \lambda})$ denotes the dual objective function, which is given by
\begin{equation}\label{equ:LG_P4}
{\mathcal G}({\bm \theta},{\bm \lambda})= f_{4a}({ {\bm \theta}})- \sum_{n=1}^N \lambda_n \left({\bm \theta}^{\rm H} {\bf e}_n {\bf e}_n^{\rm H} {\bm \theta}-1\right)
.
\end{equation}
${\mathcal G}({\bm \theta},{\bm \lambda})$ is a concave function with respect to ${\bm \theta}$.
It can be verified that, the Slater's condition is satisfied and thus the duality gap is indeed zero \cite{Boyd2004ConvexOpt}.
Then we have following lemma:

\begin{lemma}\label{Lemma_lagrange}
The optimal ${\bm \theta}$ for a given ${\bm \lambda}$ is
\begin{equation}\label{equ:opt_theta_A1}
\begin{aligned}[b]
{\bm \theta}^\circ&=
\left( \sum_{n=1}^N \lambda_n {\bf e}_n {\bf e}_n^{\rm H} +{\bf U}
\right)^{-1} {\bf v}
.
\end{aligned}
\end{equation}
The optimal dual variable vector ${\bm \lambda}^\circ$ can be determined according to the constraints in \eqref{equ:f4nconst1} via the ellipsoid method.
\end{lemma}
\begin{IEEEproof}
${\bm \theta}^\circ$ in \eqref{equ:opt_theta_A1} can be obtained by setting $\partial {\mathcal G}/\partial {\bm \theta}$ to zero.
\end{IEEEproof}

\subsubsection{Projection Step}
Denote the optimal RC for the cases when $\theta_n \in  {\mathcal F}_2$ and $\theta_n \in  {\mathcal F}_3$ by ${\theta}_n^\bullet$. Then, we have
\begin{equation}
{\theta}_n^\bullet={\rm Pj}_{\mathcal F} ({\theta}_n^\circ)
,
\end{equation}
where ${\rm Pj}_{\mathcal F}(\cdot)$ indicates the projection operation onto ${\mathcal F}$.
\begin{itemize}
\item When $\theta_n \in  {\mathcal F}_2$, the angle of $\theta_n^\bullet$ is:
\begin{equation}
\angle \theta_n^\bullet= \angle \theta_n^\circ
.
\end{equation}
\item When $\theta_n \in {\mathcal F}_3$, the angle of $\theta_n^\bullet$ is:
\begin{equation}
\angle \theta_n^\bullet = \arg \min_{\phi_n \in \{0, \frac{2 \pi}{\tau}, \cdots, \frac{2 \pi (\tau-1)}{\tau} \}} |\phi_n-\angle \theta_n^\circ|
.
\end{equation}
\end{itemize}

\subsubsection{Discussion}
Note that, the ${\bm \theta}^\bullet$ obtained by projection is not a local optimum solution of the original non-convex problem.
Thus, we only update ${\bm \theta}^\bullet$, when the constraint \eqref{equ:con_pro1} in \emph{Proposition} \ref{pro_converge} is satisfied to guarantee the convergence of {\emph{Algorithm} \ref{alg:P1}}.

In addition, another drawback of the method proposed above is the complexity.
In each iteration step of the LDD method, the highest complexity operation is to find ${\bm \theta}^\circ$ in \eqref{equ:opt_theta_A1},
in which the complexity of the summation operation $\sum_{n=1}^N \lambda_n {\bf e}_n {\bf e}_n^{\rm H} +{\bf U}$, the matrix inversion, and the final matrix multiplication are ${\mathcal O}(N)$, ${\mathcal O}(N^3)$ and ${\mathcal O}(N^2)$, respectively.
Thus, the complexity of the LDD is ${\mathcal O}(N^6)$, which is the same as the SDR technique in \cite{zhangruiIRS}.
Therefore, it is worthy designing low-complexity method to replace the conventional LDD for the NPP method.

\subsection{Iterative Reflection Coefficient Updating}\label{sec:theta_ICU}
In \cite{zhangruiIRSDiscrete}, for the single-user cases, the authors proposed an alternating optimization algorithm, which iteratively optimizes one of the $N$ RC in ${\bm \theta}$  by keeping the others fixed.
In this subsection, we extend this method to the multi-user system.
In contrast to the NPP method, the complexity of the algorithm proposed in this subsection is very low, and moreover, a local optimum can be found for $({\rm P}4a)$.

\subsubsection{Subproblem Formulation for Optimizing $\theta_n$}
Denote the element at $i$-th row and $j$-th column of ${\bm U}$ by $u_{i,j}$, and the $i$-th element of ${\bm \nu}$ by $\nu_i$.
Then, ${\bm \theta}^{\rm H} {\bm \nu}$ can be written as
\begin{equation}\label{equ:thetannu}
\begin{aligned}[b]
{\bm \theta}^{\rm H} {\bm \nu}&=\sum_{i=1}^N \theta_i^{\ast} \nu_i
\\
&=\theta_n^{\ast} \nu_n+\sum_{i=1,i \neq n}^N \theta_i^{\ast} \nu_i.
\end{aligned}
\end{equation}
Similarly, ${\bm \theta}^{\rm H}  {\bm U} {\bm \theta}$ is represented as
\begin{equation}\label{equ:thetanU}
\begin{aligned}[b]
{\bm \theta}^{\rm H}  {\bm U} {\bm \theta}&=\sum_{i=1}^N \sum_{j=1}^N \theta_i^{\ast} u_{i,j} \theta_j
\\
&=\theta_n^{\ast} u_{n,n} \theta_n+\sum_{j=1,j \neq n}^N \theta_n^{\ast} u_{n,j} \theta_j+\sum_{i=1,i \neq n}^N \theta_i^{\ast} u_{i,n} \theta_n\\
&\qquad+\sum_{i=1, i \neq n}^N \sum_{j=1, j \neq n}^N \theta_i^{\ast} u_{i,j} \theta_j.
\end{aligned}
\end{equation}
From the definition of $\bf U$ in \eqref{equ:f4U},  $\bf U$ is a hermitian matrix.
Substituting $u_{i,j}=u_{j,i}^{\ast}$ into \eqref{equ:thetanU}, we have
\begin{equation}\label{equ:thetanU2}
\begin{aligned}[b]
{\bm \theta}^{\rm H}  {\bm U} {\bm \theta}&=\sum_{i=1}^N \sum_{j=1}^N \theta_i^{\ast} u_{i,j} \theta_j
\\
&=\theta_n^{\ast} u_{n,n} \theta_n+2 {\rm Re} \left\{ \sum_{j=1,j \neq n}^N \theta_n^{\ast} u_{n,j} \theta_j \right\}\\
&\qquad+\sum_{i=1, i \neq n}^N \sum_{j=1, j \neq n}^N \theta_i^{\ast} u_{i,j} \theta_j.
\end{aligned}
\end{equation}

Substituting \eqref{equ:thetannu} and \eqref{equ:thetanU2} into \eqref{equ:f4bcr}, $f_{4a}({ {\bm \theta}})$ can be translated to a function of $\theta_n$.
After dropping all the irrelevant constant terms, we have
\begin{equation}\label{equ:f5}
\begin{aligned}[b]
f_{5}({ { \theta_n}})&=
-\theta_n^{\ast} u_{n,n} \theta_n
+2 {\rm Re} \left\{ \theta_n^{\ast} \nu_n-\sum_{j=1,j \neq n}^N \theta_n^{\ast} u_{n,j} \theta_j \right\}
\\
&= -|\theta_n|^2 A_{1,n}+ 2 {\rm Re} \left\{ {\theta}_n^{\ast} A_{2,n} \right\}
,
\end{aligned}
\end{equation}
where
\begin{align}
A_{1,n} &=u_{n,n},\\
A_{2,n}&=\nu_n-\sum_{j=1,j \neq n}^N  u_{n,j} \theta_j \label{equ:A2n}
.
\end{align}
Then, the subproblem for optimizing ${\theta_n}$ given all the other $\theta_i$ ($i \neq n$) is
\begin{align*}
({\rm P}5)\: \max_{{ {\theta_n}}} \quad &
f_{5}({ { \theta_n}})
\notag \\
{\bf s.t.} \quad
&  \theta_n \in {\mathcal F}.
\end{align*}

\subsubsection{Optimal Solutions}
$f_{5}({ { \theta_n}})$ in \eqref{equ:f5} is a concave quadratic function of $\theta_n$, and the closed-form solution for $\theta_n$ can be derived for both the convex case and the non-convex case:
\begin{itemize}
\item $\theta_n \in {\mathcal F}_1$:
In this case, $|\theta_n|^2 \leq 1$. Thus, the optimal  RC can be found by maximizing the quadratic objective, and then projecting back into a unit ball:
\begin{equation}\label{equ:thetanICU_A1}
{\theta}_n^\circ=\frac{A_{2,n}}{|A_{2,n}|} \min \left\{
1,\frac{|A_{2,n}|}{A_{1,n}}
\right\}
.
\end{equation}
\item {$\theta_n \in {\mathcal F}_2$}:
In this case, $|\theta_n|^2 = 1$. Hence, $f_{5}({ { \theta_n}})=-A_{1,n}+ 2 {\rm Re} \left\{ {\theta}_n^{\ast} A_{2,n}\right\}$, which is a linear function.
Then, the optimal RC for $({\rm P}5)$ can be obtained from
\begin{equation}
\begin{aligned}[b]
\angle {\theta}_n^\circ &=\arg \min_{\varphi_n \in [0, 2 \pi)} |\varphi_n-\angle {A}_{2,n}|\\
&=\angle {A}_{2,n}
.
\end{aligned}
\end{equation}
\item {$\theta_n \in {\mathcal F}_3$}:
In this case, we also have $|\theta_n|^2 = 1$. The optimal RC is
\begin{equation}
\begin{aligned}[b]
\angle {\theta}_n^\circ =\arg \min_{\varphi_n \in \{0, \frac{2 \pi}{\tau}, \cdots, \frac{2 \pi (\tau-1)}{\tau} \}} |\varphi_n-\angle {A}_{2,n}|
.
\end{aligned}
\end{equation}
\end{itemize}
Finally, all the reflection coefficients can be optimized based on $({\rm P}5)$ in the order from $n=1$ to $n=N$ and repeatedly.
This method is named as  the \emph{iterative reflection coefficient updating} (ICU).

\subsubsection{Discusssion}
${\theta}_n^\circ$ provided above is the optimal solution for $({\rm P}5)$, which means that we always find the optimal ${\theta}_n^\circ$ while fixing other RC values.
As a result, the ICU algorithm will converge to  a local optimum of $({\rm P}4a)$ for all the three RC assumptions.
Especially, when $\theta_n \in {\mathcal F}_1$ for all $n$, this local optimum is the global optimal solution, since $({\rm P}5)$ is convex in this case.
Therefore, in the $i$-th iteration of {\emph{Algorithm} \ref{alg:P1}}, if we initial the ICU by using ${\bm \theta}^{(i-1)}$, the updated ${ {\bm \theta}^{(i)} }$ always satisfies the constraint \eqref{equ:con_pro1} in \emph{Proposition} \ref{pro_converge}:
\begin{equation}
f_{4a}({ {\bm \theta}^{(i)} }) \geq f_{4a}( {\bm \theta}^{(i-1)}),
\end{equation}
and thus {\emph{Algorithm} \ref{alg:P1}} will converge.

From \eqref{equ:A2n}, the complexity for solving $({\rm P}5)$ is ${\mathcal O}(N)$. In every iteration step of ICU, we need to solve $N$ times $({\rm P}5)$ for all the $N$ RC values in ${\bm \theta}$. Thus the complexity of the ICU is ${\mathcal O}(N^2)$.
Since ICU can also find the optimal solution for $\theta_n \in {\mathcal F}_1$ case, it can be used to replace the LDD for the NPP method by applying ${\theta}_n^\circ$ in \eqref{equ:thetanICU_A1}. Through such operation, the complexity of NPP is reduced to ${\mathcal O}(N^2)$.

\subsection{Alternating Direction Method of Multipliers}
One drawback of the ICU algorithm is that, the RC values in ${\bm \theta}$ should be optimized one by one.
Although the complexity is low, it may still cost long time when $N$ is large.
In this section, we try to propose algorithm which can optimize ${\bm \theta}$ in parallel, meanwhile the complexity should be much lower than the NPP method.

\subsubsection{Problem Transform}
Roughly speaking, when ${\theta_n} \in {\mathcal F}_2$ or ${\theta_n} \in {\mathcal F}_3$), $({\rm P}4a)$ is a convex optimization problem with some additional non-convex constrains.
Recently, a heuristic method has been proposed for this kind of problem by employing the \emph{alternating direction method of multipliers} (ADMM) \cite{Wang2019ADMM,Boyd2016Heuristicnonconvexset}.
Although this heuristic non-convex ADMM may not find an optimal point, it can be dramatically fast to carry out a ``good'' solution \cite{Boyd2016Heuristicnonconvexset}.

Let's introduce an auxiliary vector ${\bf q}$ for ${\bm \theta}$, and a penalty term for ${\bf q}\neq{\bm \theta}$.
Then, $({\rm P}4a)$ is equivalently represented as
\begin{subequations}
\begin{align}
({\rm P}4c)\: \max_{ {\bm \theta}, {\bf q} } \quad &
 f_{4a}({\bf q})-\frac{\mu}{2} \|{\bf q}-{\bm \theta}\|^2_2
\notag \\
{\bf s.t.} \quad & {\bf q}={\bm \theta},\label{equ:P4cc1} \\
& \theta_n \in {\mathcal F}, \quad \forall n=1,\cdots,N, \label{equ:P4cc2}
\end{align}
\end{subequations}
where $\mu>0$ is the penalty parameter.
We need two Lagrange variables ${\bm \lambda}_{\rm R}=[\lambda_{{\rm R},1},\cdots,\lambda_{{\rm R},N}]^{\rm T}$ and ${\bm \lambda}_{\rm I}=[\lambda_{{\rm I},1},\cdots,\lambda_{{\rm I},N}]^{\rm T}$ for
${\rm Re} \{{\bf q}-{\bm \theta}\}=0$ and ${\rm Im} \{{\bf q}-{\bm \theta}\}=0$, respectively, since the constraint in \eqref{equ:P4cc1} is a complex equation.
Then, the  Lagrangian of ({\rm P}4c) is:
\begin{equation}\label{equ:P4cLagran}
\begin{aligned}
{\mathcal G}({\bf q}, {\bm \theta},{\bm \lambda}_{\rm R},{\bm \lambda}_{\rm I})
&=-{\bm q}^{\rm H}  {\bm U} {\bm q}
- \sum_{n=1}^N {\mathbbm{1}}_{{\mathcal F}}(\theta_n)-\frac{\mu}{2} \|{\bf q}-{\bm \theta}\|^2_2
\\
& \quad+  {\rm Re}\left\{ 2 {\bm q}^{\rm H} {\bm \nu}+
    \left({\bm \lambda}_{\rm R}+j {\bm \lambda}_{\rm I}\right)^{\rm H} \left({\bf q}-{\bm \theta} \right)
    \right\}
,
\end{aligned}
\end{equation}
where ${\mathbbm{1}}_{{\mathcal F}}(\cdot)$ is the indicator function of set $\mathcal F$ (i.e., ${\mathbbm{1}}_{{\mathcal F}}(\theta_n)=0$ if $\theta_n \in {\mathcal F}$; otherwise, equals infinity).
Thus the  dual problem is formulated as
\begin{align*}
({\rm P}6)\: \min_{ {\bm \lambda}_{\rm R},{\bm \lambda}_{\rm I} } \quad &
{\mathcal L} ({\bm \lambda}_{\rm R},{\bm \lambda}_{\rm I})=\max_{{ {\bm \theta}, {\bm q}}} \left\{
 {\mathcal G}({\bf q}, {\bm \theta},{\bm \lambda}_{\rm R},{\bm \lambda}_{\rm I})
\right\}.
\end{align*}

It can be verified that, the Slater's condition \cite{Boyd2004ConvexOpt} is satisfied when ${\mathcal F}={\mathcal F}_1$. Thus the duality gap is indeed zero, i.e., $({\rm P}6)$ is equivalent to $({\rm P}4c)$.
However, when ${\mathcal F}={\mathcal F}_2$ or ${\mathcal F}={\mathcal F}_3$, $({\rm P}4c)$ is non-convex and the duality gap exists. In these cases,
the optimal objective value of $({\rm P}6)$ only serves an upper bound of the primal problem $({\rm P}4c)$.
The benefit of this transformation is that, solving $({\rm P}6)$ is relatively simpler than solving the primal problem $({\rm P}4c)$.

\begin{table*}[!t]
\footnotesize
\renewcommand{\arraystretch}{1.3}
\caption{Comparison of the Three RC Optimization Algorithms for $({\rm P}4)$}
\label{tablepm}
\centering
\begin{tabular}{c|c|c|c|c}
\hline
Algorithm & Complexity & Implementation & \tabincell{c}{Convergence \\(${\mathcal F}_1$)}
& \tabincell{c}{Convergence \\(${\mathcal F}_2$, ${\mathcal F}_3$)}  \\
\hline
\tabincell{c}{NPP \\(LDD based)}  & ${\mathcal O}(N^6)$ & Parallel & Optimal & N/A   \\
\hline
ICU  & ${\mathcal O}(N^2)$ & Serial & Optimal & {Local optimum}   \\
\hline
ADMM  & ${\mathcal O}(N^3)$ & Parallel & Optimal & {Optimal for dual problem}   \\
\hline
\end{tabular}
\label{tablec}
\end{table*}

\subsubsection{ADMM for $({\rm P}6)$}
In this part, we solve $({\rm P}6)$ via ADMM which has the following iterative form:
\begin{align}
{\bm \theta}^{t+1}&= \arg \max_{{\bm \theta}} {\mathcal G}({\bf q}^{t}, {\bm \theta},{\bar {\bm \lambda}}^{t}), \label{equ:Admm_Theta}\\
{\bf q}^{t+1}&= \arg \max_{{\bf q}} {\mathcal G}({\bf q}, {\bm \theta}^{t+1},{\bar {\bm \lambda}}^{t}), \label{equ:Admm_Q}\\
{\bar {\bm \lambda}}^{t+1}
&= {\bar {\bm \lambda}}^{t}- \mu \left({\bf q}^{t+1}-{\bm \theta}^{t+1}\right), \label{equ:Admm_lambda}
\end{align}
where ${\bar {\bm \lambda}}={\bm \lambda}_{\rm R}+j {\bm \lambda}_{\rm I}$, and $t$ is the iteration index.

To be specific,
in \eqref{equ:Admm_Theta}, ${\bm \theta}^{t+1}$  is optimized given ${\bf q}^{t}$ and ${\bar {\bm \lambda}}^{t}$.
The optimal ${\bm \theta}$  is
\begin{equation}\label{equ:optthetau_P5}
 {\bm \theta}^{t+1}={\rm Pj}_{\mathcal F} \left( {\bf q}^{t}-\frac{1}{\mu} {\bar {\bm \lambda}}^{t} \right)
.
\end{equation}
Let ${\bar {\bm \theta}}={\bf q}^{t}-\frac{1}{\mu} {\bar {\bm \lambda}}^{t}$. The projection operation in \eqref{equ:optthetau_P5} is:
\begin{itemize}
\item When $\theta_n \in {\mathcal F}_1$:
\begin{equation}
\theta_n^{t+1}=\frac{{\bar \theta}_n}{|{\bar \theta}_n|} \min \left\{1, |{\bar \theta}_n|\right\};
\end{equation}
\item When $\theta_n \in {\mathcal F}_2$:
\begin{equation}
\angle  \theta_n^{t+1}=\angle {\bar \theta}_n
;
\end{equation}
\item When $\theta_n \in {\mathcal F}_3$:
\begin{equation}
\angle  \theta_n^{t+1}=\arg \min_{\varphi_n \in \{0, \frac{2 \pi}{\tau}, \cdots, \frac{2 \pi (\tau-1)}{\tau} \}} |\varphi_n-\angle {\bar \theta}_n|
.
\end{equation}
\end{itemize}

Then, in \eqref{equ:Admm_Q}, ${\bf q}$ is optimized given ${\bm \theta}^{t+1}$ and ${\bar {\bm \lambda}}^{t}$, and we have
\begin{equation}\label{equ:optq_P5}
{\bf q}^{t+1}=\left(2 {\bf U}+ \mu {\bf I}_N \right)^{-1} \left(2{\bf v}+{\bar {\bm \lambda}}^{t}+\mu {\bm \theta}^{t+1} \right)
.
\end{equation}
In the end, the  Lagrange variables ${\bar {\bm \lambda}}^{t+1}$ is updated in \eqref{equ:Admm_lambda}.

Note that, when ${\mathcal F}={\mathcal F}_1$, the ADMM algorithm will converge to the global optimum \cite{Boyd2011ADMM}.
However, this is not necessarily true when  ${\mathcal F}={\mathcal F}_2$ or ${\mathcal F}={\mathcal F}_3$, since  $({\rm P}4c)$ is a non-convex optimization problem.
For these cases, the convergence condition of the ADMM algorithm is presented in the following lemma.

\begin{lemma}\label{Lemma_ADMM}
When $\theta_n$ is belongs to set ${\mathcal F}_2$ or ${\mathcal F}_3$, the ADMM algorithm presented above guarantees to converge, if the  penalty parameter $\mu$ satisfies:
\begin{equation}\label{equ:ADMM_conv}
\begin{aligned}[b]
\frac{\mu}{2} {\bf I}_N - {\bm U} \succ 0
.
\end{aligned}
\end{equation}
\end{lemma}

\begin{IEEEproof}
Please see the detailed proof in Appendix \ref{ADMM_app}.
\end{IEEEproof}

In this paper, we choose $\mu=\iota \|{\bm U}\|_2$, where $\iota\geq 1$ is the minimum integer which satisfies \eqref{equ:ADMM_conv}. 

\subsubsection{Disucssion}
From \emph{Lemma} \ref{Lemma_ADMM}, we only know that the ADMM algorithm converge. But it need not be to a global or even local optimal. Thus, we need to check whether the output ${\bm \theta}$ satisfies the the constraint \eqref{equ:con_pro1} to guarantee the convergence of {\emph{Algorithm} \ref{alg:P1}}.

It is seen that, the highest complexity operation in the ADMM algorithm is to update ${\bf q}^{t+1}$ in \eqref{equ:optq_P5}.
Contrary to the LDD, the matrix inverse term in \eqref{equ:optq_P5} is a constant in every iteration step, and thus it can be pre-computed in the initialization step.
Therefore, the complexity of the ADMM algorithm is  ${\mathcal O}(N^3)$, which is much lower than the LDD in Section \ref{theta_LDD} and the SDR technique in \cite{zhangruiIRS}.
In addition, although the complexity of ADMM is slightly higher than the ICU in Section \ref{sec:theta_ICU}, the RC vector $\bm \theta$ is updated in parallel instead of the serial operation in ICU. Hence, the ADMM algorithm may converge faster via parallel computing, especially when $N$ is large.
Finally, we summarize the comparison of the three RC optimization algorithms in {\tablename~\ref{tablepm}}.
Note that, both the ICU and ADMM algorithms can be employed to replace the LDD in NPP method, and then the complexity of NPP is reduced to ${\mathcal O}(N^2)$ and ${\mathcal O}(N^3)$, respectively.

\section{Numerical Results}\label{simulation}
\subsection{Simulation Scenario}\label{simulation_link}
\begin{figure}
[!t]
\centering
\includegraphics[width=.8\columnwidth]{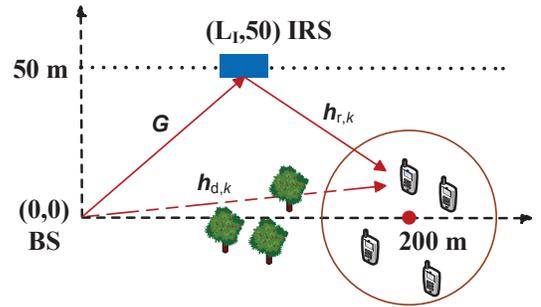}
\caption{The simulated IRS-aided $K$-user MISO communication scenario.}
\label{simulation_scena}
\end{figure}

In this section, numerical examples are provided to validate the effectiveness of the proposed algorithms.
We consider an IRS-aided femtocell network illustrated in {\figurename~\ref{simulation_scena}}, in which the BS and IRS are located at $(0,0)$ and ($L_{\rm I},50$ m), respectively.
The BS is equipped with $4$ antennas ($M=4$), and the reflection efficiency of IRS is set as $\eta=0.8$.
There are $4$ users ($K=4$) uniformly and randomly distributed in a circle centered at ($200$ m$,0$) with radius $10$ m.
The transmission bandwidth is $200$ kHz, and the noise power spectral density is $-170$ dBm/Hz. Thus  the background noise at the receivers is $\sigma_0^2=-117$ dBm.

The large-scale fading of the direct channel (with respect to the power of ${\bf h}_{{\rm d},k}$) is modelled as
\begin{equation}\label{equ:Direct_link_passloss}
\begin{aligned}[b]
\kappa_{{\rm D},k}= C_{\rm p} \varsigma_{\rm B} \varsigma_{k} d_{{\rm D},k}^{- {\varrho_{\rm D}}},
\end{aligned}
\end{equation}
where $d_{{\rm D},k}$ denotes the link distance between the BS and the $k$-th user, ${\varrho_{\rm D}}=3.5$ is the path loss exponent, $C_{\rm p}$ is a constant with respect to the wavelength, and $\varsigma_{\rm B}$ and $\varsigma_{k}$ denote the antenna gain of the BS and $k$-th user, respectively.
We assume that $C_{\rm p} \varsigma_{\rm B} \varsigma_{k}=-30$  dB, which indeed is the path loss at the reference distance ($d_{{\rm D},k}=1$ m).
Besides, we assume that the IRS-aided link experiences the free-space path loss, since the location of IRS is usually carefully chosen.
Then, the large-scale fading with respect to the power of ${\bf G}$ and ${\bf h}_{{\rm r},k}$ are denoted by $\kappa_{{\rm G}}=C_{\rm p} \varsigma_{\rm B} \varsigma_{\rm I} d_{{\rm G}}^{-{\varrho_{\rm I}}}$ and $\kappa_{{\rm r},k}=C_{\rm p} \varsigma_{\rm I} \varsigma_{k} d_{{\rm r},k}^{-{\varrho_{\rm I}}}$, respectively, where ${\varrho_{\rm I}}=2$ is the path loss exponent, $d_{{\rm G}}$ and $d_{{\rm r},k}$ are the distance from BS to IRS and from IRS to the $k$-th user, respectively, and $\varsigma_{\rm I}$ is the reflection gain of the IRS element.
Since the IRS-aided link is the concatenation of ${\bf G}$ and ${\bf h}_{{\rm r},k}$, the total path loss is
\begin{equation}\label{equ:IRS_link_passloss}
\begin{aligned}[b]
\kappa_{{\rm I},k}&=\kappa_{{\rm G}} \kappa_{{\rm r},k}\\
&=C_{\rm p}^2 \varsigma_{\rm B} \varsigma_{k} \varsigma_{\rm I}^2 \left(d_{{\rm G}} d_{{\rm r},k}\right)^{-{\varrho_{\rm I}}}
.
\end{aligned}
\end{equation}
Comparing \eqref{equ:IRS_link_passloss} with \eqref{equ:Direct_link_passloss},
one can see that the signal reflected by the IRS suffered from the ``double-fading'' effect \cite{Griffin2009LinkBudget}.
Nevertheless, the reflection gain of the IRS elements is usually much higher than the antenna gain of the mobile station  thanks to the recent advances in meta-materials.
Denote by the relative reflection gain $\xi=\frac{\varsigma_{\rm I}}{\sqrt{ \varsigma_{\rm B} \varsigma_{k} }}$.
Then, in the simulation scenario shown in {\figurename~\ref{simulation_scena}}, the direct-link path loss from BS to ($200$ m$,0$) is about -111 dB.
If we have $\xi=10$ dB, the path loss of the IRS-aided link is about -122 dB, and in this case the IRS may potentially double the average receive power at the user side using $N=10$ elements.

For simplicity, we assume the Rayleigh fading model to account for small-scale fading.
The weights $\omega_k$ are set to be equal in all the simulations.
All the simulation results are obtained by averaging over $10^4$ channel realizations.
Specifically, we first generate $100$ snapshots, in which the locations of the mobile users are randomly chosen.
Then, for each snapshot, we further generate $100$ channel realizations with independent small-scale fading.

\subsection{Benchmarks and Initialization}
We compare the performance of the proposed algorithms with the following $2$ baselines:
\begin{itemize}
\item {\bf Baseline 1} (Without the aid of IRS): Let $N=0$, and then the active beamforming is optimized via  the WMMSE in \cite{WMMSE}. In particular, this baseline can be obtained by skipping the {\emph {Step}} 3.1 and 3.2 of the {\emph{Algorithm} \ref{alg:P1}} \cite{YuWei2018FP2}.
\item {\bf Baseline 2} (Random passive beamforming): The  phase-shift matrix of the IRS is not optimized, in which the RCs are chosen by random values from set ${\mathcal F}_2$. Then, WMMSE is adopted to optimize the active beamforming at the BS.
\end{itemize}

Besides, the WSR maximization problem $({\rm P}1)$ investigated in this paper is non-convex, and the proposed {\emph{Algorithm} \ref{alg:P1}} only finds a suboptimal solution. Therefore, the performance of {\emph{Algorithm} \ref{alg:P1}} is sensitive to the initialization of ${ {\bm W}}$ and ${\bm \Theta}$. In this paper, for the three different assumptions of RC values, we employ different initializations:
\begin{itemize}
\item $\theta \in {\mathcal F}_1$: In this case, $\theta$ in ${\bm \Theta}$ is initialized by random values in ${\mathcal F}_2$, and ${ {\bm W}}$ is initialized by the zero-forcing beamforming.
\item $\theta \in {\mathcal F}_2$ or $\theta \in {\mathcal F}_3$: In this case, the constraint sets of $\theta$ is non-convex, and the proposed algorithm is more vulnerable to be trapped in local optimum. Hence, we initialize  ${ {\bm W}}$ and ${\bm \Theta}$ by the solutions of the $\theta \in {\mathcal F}_1$ (convex constraint set) case, in which both the active and passive beamforming have found good directions.
\end{itemize}

\subsection{Sum Rate versus Transmit Power $P_{\rm T}$}
\begin{figure}
[!t]
  \centering
  \subfigure[$P_{\rm T}$ vs $R$]{
    \label{wsr_vs_PT:a} 
    \includegraphics[width=1\columnwidth]{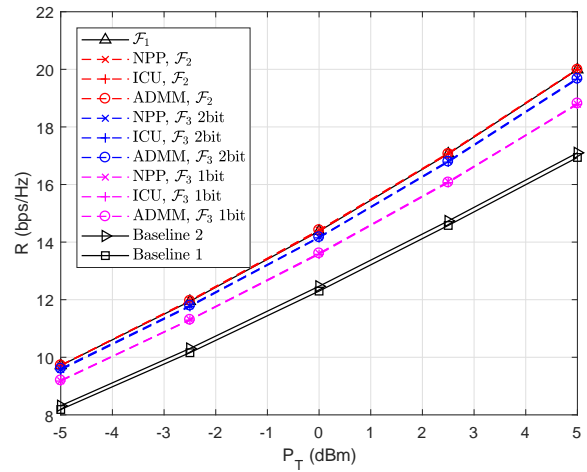}}
  \subfigure[$P_{\rm T}=0$ dBm]{
    \label{wsr_vs_PT:b} 
    \includegraphics[width=1\columnwidth]{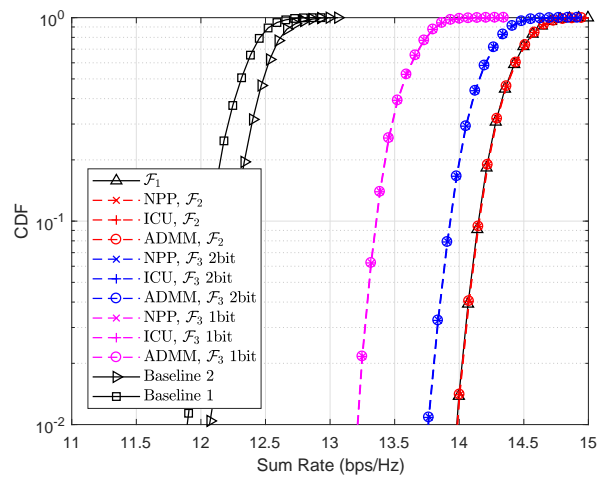}}
  \caption{The sum rate versus transmit power, when $N=10$ and $\xi=10$ dB.}
  \label{wsr_vs_PT} 
\end{figure}

{\figurename~\ref{wsr_vs_PT:a}} illustrate the average sum rate of different schemes with respect to the transmit power $P_{\rm T}$, when $N=10$ and $\xi=10$ dB.
The average sum rate over different channel realizations is denoted by $R$.
We set $L_{\rm I}=100$ m, and thus the IRS is deployed at ($100$ m, $50$ m).
It is seen that, the performance gain of the random passive beamforming scheme (Baseline 2) is very small, since most reflected signals cannot arrive the receivers of the mobile users.
On the other hand, significant performance gains are achieved by joint active and passive beamforming optimization, and all the three proposed algorithms have almost the same performance.
In particular, the joint beamforming schemes achieve about $3$ dB gain comparing with Baseline 1 as expected.
The performance loss is negligible when the ideal RC constraint reduces to  $\theta \in {\mathcal F}_2$.
This reveal that, although we relax the constraint of $\theta$ from  $|\theta|=1$ to $|\theta|\leq 1$, the amplitude of the optimal $\theta$ is still very close to $1$.
In addition, we also observe that the  ``1-bit'' phase shifter still achieves about $1.5$ dB gain, and the  ``2-bit'' phase shifter may obtain almost the full beamforming gain compared with the continuous phase shifter cases.

Next, in {\figurename~\ref{wsr_vs_PT:b}}, we fix the transmit power $P_{\rm T}=0$ dBm, and plot the \emph{cumulative distribution function} (CDF) of the sum rate over different snapshots.
It is seen that, the performance gains of all the proposed schemes under different RC assumptions are stable over the CDF curves, and also keep consistent with their counterparts in {\figurename~\ref{wsr_vs_PT:a}}.
Therefore, we conclude that, with high probability, the performance of the proposed algorithms will be good irrespective of user location.

\subsection{IRS Size and Material}
\begin{figure}
[!t]
  \centering
  \subfigure[$R$ vs $N$, when $P_{\rm T}=0$ dBm and $\xi=10$ dB.]{
    \label{N_vs_PT_fixeta} 
    \includegraphics[width=1\columnwidth]{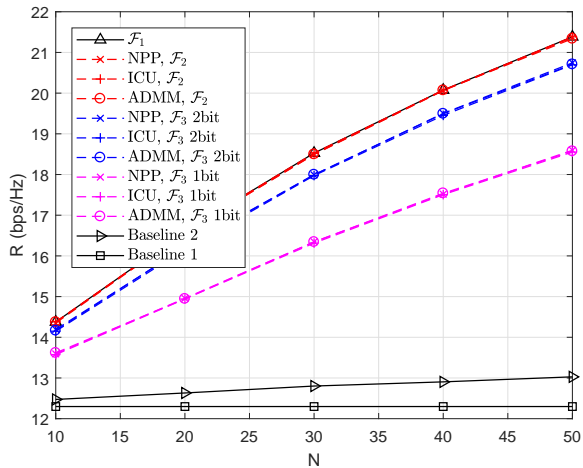}}
  \subfigure[$R$ vs $\xi$, when $P_{\rm T}=-5$ dBm and $N=10$.]{
    \label{rate_vs_xi} 
    \includegraphics[width=1\columnwidth]{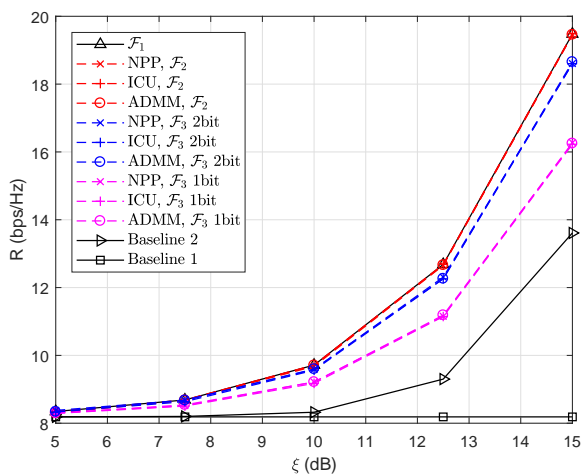}}
  \caption{The sum rate versus IRS size $N$ and relative reflection gain $\xi$.}
  \label{rate_vs_N_xi} 
\end{figure}

{\figurename~\ref{N_vs_PT_fixeta}} compares the average sum rate with the size $N$ of IRS, while the transmit power of BS is fixed to $0$ dBm and the relative reflection gain of IRS is $\xi=10$ dB.
It is observed that the performance of all the schemes with the aid of IRS increases with the increase of $N$, since the sum power of the signals reflected by the IRS  becomes stronger.
However, the quantization loss of the discrete phase shifter also increases as $N$ increases.
Hence, we prefer high-order quantization for cases with large $N$.
In addition, we observe that, to achieve $R=20$ bps/Hz, the size should be increased from $10$ to $40$ (for continuous-phase-shifter cases), i.e., $6$ dB.
On the other hand, $R=20$ bps/Hz can also be achieved by increasing $P_{\rm T}$ from $0$ dBm to $5$ dBm (only $5$ dB) as shown in {\figurename~\ref{wsr_vs_PT:a}}.
This is because, when $N$ increases, only the IRS-assisted link is enhanced, while when the transmit power of BS increases, both the direct link and the IRS-assisted link get  benefits.

%

Then, we investigate the impact of the relative reflection gain $\xi$ on the average sum rate.
It is known from \cite{LiuFu2019Metasurface} that, $\xi$ is dominated by the resistance of the integrated circuits in the IRS element, and recent research has shown that $\xi$ can be greatly improved by exploiting the negative resistance materials \cite{Kimionis2014enhanceBackefficiency}.\footnote{The negative resistance materials are generally comprising of active components. Thus in this case, the IRS becomes semi-passive.}
{\figurename~\ref{rate_vs_xi}} illustrates the average sum rate of different schemes with respect to $\xi$, when $P_{\rm T}=-5$ dBm and $N=10$.
One can observe that, the IRS-aided system with continuous phase shifter may achieve about $R=20$ bps/Hz by increasing $\xi$ from $10$ dB to $15$ dB. 
Comparing with {\figurename~\ref{wsr_vs_PT:a}} and {\figurename~\ref{N_vs_PT_fixeta}}, we conclude that, to increase $\xi$ is much more effective than to increase $P_{\rm T}$ and $N$ for improving $R$.
This is because, the reflection gain of IRS is counted twice during the reflecting operation according to \eqref{equ:IRS_link_passloss}.
Moreover, when $\xi$ is large, even the random passive beamforming scheme (Baseline 2) could achieve a significant rate gain.
Therefore, it is very attractive to investigate how to improve $\xi$ for the IRS elements with new reflection materials.

\subsection{Deployment Location}
\begin{figure}
[!t]
\centering
\includegraphics[width=1\columnwidth]{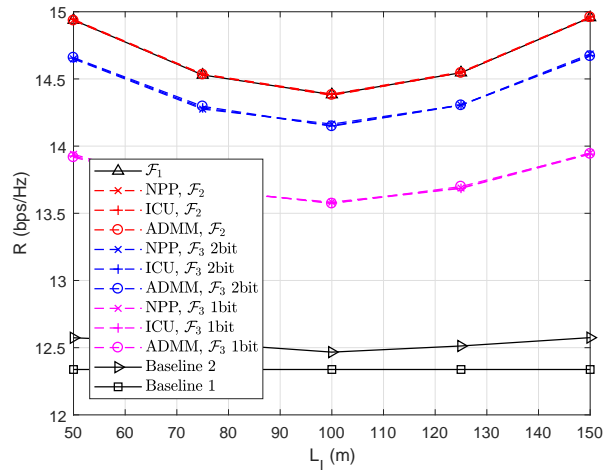}
\caption{The sum rate versus the location of IRS, where IRS are located at ($L_{\rm I},50$ m), and we set $P_{\rm T}=0$ dBm, $N=10$, and $\xi=10$ dB.}
\label{location_vs_R}
\end{figure}

Finally, we discuss on the impact of the IRS deployment locations.
We move the IRS from $L_{\rm I}=50$ m to $L_{\rm I}=200$ m, and plot the average sum rate of different schemes with respect to $L_{\rm I}$ in {\figurename~\ref{location_vs_R}}, while setting $P_{\rm T}=0$ dBm, $N=10$, and $\xi=10$ dB.
It can be seen, the performance gain of the IRS-aided system increases when the IRS is deployed closer to the BS or the cluster of users,
and deploying the IRS at the center place ($L_{\rm I}=100$ m) is the worst case.
This conclusion can also be inferred from the double-fading path-loss model in \eqref{equ:IRS_link_passloss}.
However, when the IRS is deployed too close to the BS or users, the  propagation condition may get as worse as the direct link.
Thus there exists a trade-off between the propagation condition and the double-fading effect.
In practice, for the convenience of controlling between BS and IRS, the IRS is preferred to be deployed close to the BS, while guaranteeing  high-quality BS-IRS-user links at the same time.

\section{Conclusion}\label{conclusion}
In this paper, we investigate the  IRS-aided multiuser downlink MISO system.
Specifically, a joint active and passive beamforming problem is formulated to maximize the WSR under the BS transmit power constraint.
To tackle this non-convex problem, an iterative method has been developed by utilizing the recently proposed  fractional programming technique.
In addition, three low-complexity algorithms are proposed to solve the passive beamforming problem with closed-form solutions.
All the three algorithms are applicable not only to the continuous phase-shift IRS but also the discrete  phase-shift IRS.
Extensive simulation results demonstrated that the proposed joint beamforming scheme achieves significant capacity gain compared with the conventional system without the IRS and the IRS-aided system employing random passive beamforming.
Moreover, it is also shown that the IRS with  2-bit quantizer may achieve sufficient capacity gain with only a small performance degradation.

\appendices
\section{Proof of Lemma \ref{Lemma_ADMM}}\label{ADMM_app}
From \eqref{equ:optq_P5}, we have following equation:
\begin{equation}\label{equ:Proof_L3_qlambda}
{\bar {\bm \lambda}}^{t}+\mu {\bm \theta}^{t+1} =
\left(2 {\bf U}+ \mu {\bf I}_N \right){\bf q}^{t+1}-2{\bf v}
.
\end{equation}
Substituting \eqref{equ:Proof_L3_qlambda} into \eqref{equ:Admm_lambda}, we have
\begin{equation}\label{equ:Proof_L3_lambda}
\begin{aligned}[b]
{\bar {\bm \lambda}}^{t+1}
&= {\bar {\bm \lambda}}^{t}- \mu \left({\bf q}^{t+1}-{\bm \theta}^{t+1}\right) \\
&=\left(2 {\bf U}+ \mu {\bf I}_N \right){\bf q}^{t+1}-2{\bf v}- \mu{\bf q}^{t+1}\\
&=2 {\bf U} {\bf q}^{t+1}-2{\bf v}
.
\end{aligned}
\end{equation}
Therefore, ${\bar {\bm \lambda}}$ can be replaced by a function of ${\bf q}$:
\begin{equation}\label{equ:Proof_L3_lambdaq}
{\bar {\bm \lambda}}=2 {\bf U} {\bf q}-2{\bf v}
.
\end{equation}
By substituting \eqref{equ:Proof_L3_lambdaq} into ${\mathcal G}({\bf q}, {\bm \theta},{\bm \lambda}_{\rm R},{\bm \lambda}_{\rm I})$, we get a new function:
\begin{equation}\label{equ:P4cLagran_L3}
\begin{aligned}[b]
{\mathcal V}({\bf q}, {\bm \theta})
&=-{\bm q}^{\rm H}  {\bm U} {\bm q}
- \sum_{n=1}^N {\mathbbm{1}}_{{\mathcal F}}(\theta_n)-\frac{\mu}{2} \|{\bf q}-{\bm \theta}\|^2_2
\\
& \qquad+  {\rm Re}\left\{ 2 {\bm q}^{\rm H} {\bm \nu}+
    \left(2 {\bf U} {\bf q}-2{\bf v} \right)^{\rm H} \left({\bf q}-{\bm \theta} \right)
    \right\}\\
&=- {\bm q}^{\rm H}  (\frac{\mu}{2} {\bf I}_N - {\bm U}) {\bm q}-\frac{\mu}{2} {\bm \theta}^{\rm H} {\bm \theta}
- \sum_{n=1}^N {\mathbbm{1}}_{{\mathcal F}}(\theta_n)\\
& \qquad+{\rm Re}\left\{ 2{\bf v}^{\rm H} {\bm \theta}-2 {\bm q}^{\rm H} {\bm U} {\bm \theta} + \mu {\bm \theta}^{\rm H} {\bm q}
\right\}
.
\end{aligned}
\end{equation}
It is easy to verify that, when $\frac{\mu}{2} {\bf I}_N - {\bm U} \succ 0$ in \eqref{equ:ADMM_conv} is satisfied,
the ADMM iteration from  \eqref{equ:Admm_Theta} to \eqref{equ:Admm_lambda} is equivalent to the following coordinate ascent iteration to ${\mathcal V}({\bf q}, {\bm \theta})$:
\begin{align}
{\bm \theta}^{t+1}&= \arg \max_{{\bm \theta}} {\mathcal V}({\bf q}^{t}, {\bm \theta}) \\
{\bf q}^{t+1}&= \arg \max_{{\bf q}} {\mathcal V}({\bf q}, {\bm \theta}^{t+1})
.
\end{align}
Hence, the ADMM algorithm guarantees to converge.

\bibliographystyle{IEEEtran}
\bibliography{IEEEabrv,mybib_draft2}

\begin{thebibliography}{10}
\providecommand{\url}[1]{#1}
\csname url@samestyle\endcsname
\providecommand{\newblock}{\relax}
\providecommand{\bibinfo}[2]{#2}
\providecommand{\BIBentrySTDinterwordspacing}{\spaceskip=0pt\relax}
\providecommand{\BIBentryALTinterwordstretchfactor}{4}
\providecommand{\BIBentryALTinterwordspacing}{\spaceskip=\fontdimen2\font plus
\BIBentryALTinterwordstretchfactor\fontdimen3\font minus
  \fontdimen4\font\relax}
\providecommand{\BIBforeignlanguage}[2]{{%
\expandafter\ifx\csname l@#1\endcsname\relax
\typeout{** WARNING: IEEEtran.bst: No hyphenation pattern has been}%
\typeout{** loaded for the language `#1'. Using the pattern for}%
\typeout{** the default language instead.}%
\else
\language=\csname l@#1\endcsname
\fi
#2}}
\providecommand{\BIBdecl}{\relax}
\BIBdecl

\bibitem{Tan2018SRA}
X.~Tan, Z.~Sun, D.~Koutsonikolas, and J.~M. Jornet, ``Enabling indoor mobile
  millimeter-wave networks based on smart reflect-arrays,'' in \emph{Proc. IEEE
  INFOCOM}, Apr. 2018, pp. 270--278.

\bibitem{Liu2019metasurface}
F.~Liu, O.~Tsilipakos, A.~Pitilakis, A.~C. Tasolamprou, M.~S. Mirmoosa, N.~V.
  Kantartzis, D.-H. Kwon, M.~Kafesaki, C.~M. Soukoulis, and S.~A. Tretyakov,
  ``Intelligent metasurfaces with continuously tunable local surface impedance
  for multiple reconfigurable functions,'' \emph{Physical Review Applied},
  vol.~11, no.~4, p. 044024, 2019.

\bibitem{cuiTJ2017metasurface}
L.~Li, T.~J. Cui, W.~Ji, S.~Liu, J.~Ding, X.~Wan, Y.~B. Li, M.~Jiang, C.-W.
  Qiu, and S.~Zhang, ``Electromagnetic reprogrammable coding-metasurface
  holograms,'' \emph{Nature Commun.}, vol.~8, no.~1, p. 197, 2017.

\bibitem{Larsson2014FDrelay}
H.~Q. Ngo, H.~A. Suraweera, M.~Matthaiou, and E.~G. Larsson, ``Multipair
  full-duplex relaying with massive arrays and linear processing,'' \emph{IEEE
  J. Sel. Areas Commun.}, vol.~32, no.~9, pp. 1721--1737, 2014.

\bibitem{Liaskos2018magzineIRS}
C.~Liaskos, S.~Nie, A.~Tsioliaridou, A.~Pitsillides, S.~Ioannidis, and
  I.~Akyildiz, ``A new wireless communication paradigm through
  software-controlled metasurfaces,'' \emph{IEEE Commun. Mag.}, vol.~56, no.~9,
  pp. 162--169, 2018.

\bibitem{Renzo2019position}
M.~Di~Renzo, M.~Debbah, D.-T. Phan-Huy, A.~Zappone, M.-S. Alouini, C.~Yuen,
  V.~Sciancalepore, G.~C. Alexandropoulos, J.~Hoydis, and H.~Gacanin, ``Smart
  radio environments empowered by {AI} reconfigurable meta-surfaces: An idea
  whose time has come,'' \emph{arXiv preprint arXiv:1903.08925}, 2019.

\bibitem{Hum2014ReflectarrayReview}
S.~V. Hum and J.~Perruisseau-Carrier, ``Reconfigurable reflectarrays and array
  lenses for dynamic antenna beam control: A review,'' \emph{IEEE Trans.
  Antennas Propag.}, vol.~62, no.~1, pp. 183--198, 2014.

\bibitem{chenjie2019IRS}
J.~Chen, Y.-C. Liang, Y.~Pei, and H.~Guo, ``Intelligent reflecting surface: A
  programmable wireless environment for physical layer security,'' \emph{arXiv
  preprint arXiv:1905.03689}, 2019.

\bibitem{Ayach2014sparseprecoding}
O.~El~Ayach, S.~Rajagopal, S.~Abu-Surra, Z.~Pi, and R.~W. Heath, ``Spatially
  sparse precoding in millimeter wave {MIMO} systems,'' \emph{IEEE Trans.
  Wireless Commun.}, vol.~13, no.~3, pp. 1499--1513, 2014.

\bibitem{AnLiu2014HybridBeam}
A.~Liu and V.~Lau, ``Phase only {RF} precoding for massive {MIMO} systems with
  limited {RF} chains,'' \emph{IEEE Trans. Signal Process.}, vol.~62, no.~17,
  pp. 4505--4515, 2014.

\bibitem{YuWei2016HybridBeam}
F.~Sohrabi and W.~Yu, ``Hybrid digital and analog beamforming design for
  large-scale antenna arrays,'' \emph{IEEE J. Sel. Topics Signal Process.},
  vol.~10, no.~3, pp. 501--513, 2016.

\bibitem{Larsson2013ConstprocodingMU}
S.~K. Mohammed and E.~G. Larsson, ``Per-antenna constant envelope precoding for
  large multi-user {MIMO} systems,'' \emph{IEEE Trans. Commun.}, vol.~61,
  no.~3, pp. 1059--1071, 2013.

\bibitem{Larsson2013ConstprocodingFS}
------, ``Constant-envelope multi-user precoding for frequency-selective
  massive {MIMO} systems,'' \emph{IEEE Wireless Commun. Lett.}, vol.~2, no.~5,
  pp. 547--550, 2013.

\bibitem{Larsson2012SingleuserbeamCEC}
------, ``Single-user beamforming in large-scale {MISO} systems with
  per-antenna constant-envelope constraints: The doughnut channel,'' \emph{IEEE
  Trans. Wireless Commun.}, vol.~11, no.~11, pp. 3992--4005, 2012.

\bibitem{zhangrui2018GcomIRS}
Q.~Wu and R.~Zhang, ``Intelligent reflecting surface enhanced wireless network:
  Joint active and passive beamforming design,'' in \emph{Proc. IEEE Globecom},
  Dec. 2018, pp. 1--6.

\bibitem{zhangruiIRS}
------, ``Intelligent reflecting surface enhanced wireless network via joint
  active and passive beamforming,'' \emph{arXiv:1809.01423}, 2018.

\bibitem{Yuen2018ZF}
C.~Huang, A.~Zappone, M.~Debbah, and C.~Yuen, ``Achievable rate maximization by
  passive intelligent mirrors,'' in \emph{Proc. IEEE ICASSP}, May. 2018, pp.
  3714--3718.

\bibitem{YuenChauIRS}
C.~Huang, A.~Zappone, G.~C. Alexandropoulos, M.~Debbah, and C.~Yuen, ``Large
  intelligent surfaces for energy efficiency in wireless communication,''
  \emph{arXiv:1810.06934}, 2018.

\bibitem{Dennah2019IRS}
Q.-U.-A. Nadeem, A.~Kammoun, A.~Chaaban, M.~Debbah, and M.-S. Alouini, ``Large
  intelligent surface assisted {MIMO} communications,''
  \emph{arXiv:1903.08127}, 2019.

\bibitem{Yuen2018EEIRSLowbit}
C.~Huang, G.~C. Alexandropoulos, A.~Zappone, M.~Debbah, and C.~Yuen, ``Energy
  efficient multi-user {MISO} communication using low resolution large
  intelligent surfaces,'' in \emph{Proc. IEEE Globecom Workshops}, Dec. 2018,
  pp. 1--6.

\bibitem{zhangruiIRSDiscrete}
Q.~Wu and R.~Zhang, ``Beamforming optimization for intelligent reflecting
  surface with discrete phase shifts,'' \emph{arXiv:1810.10718}, 2018.

\bibitem{YuWei2018FP1}
K.~Shen and W.~Yu, ``Fractional programming for communication systems--{Part
  I}: Power control and beamforming,'' \emph{IEEE Trans. Signal Process.},
  vol.~66, no.~10, pp. 2616--2630, 2018.

\bibitem{Liu2013AmBC}
V.~Liu, A.~Parks, V.~Talla, S.~Gollakota, D.~Wetherall, and J.~R. Smith,
  ``Ambient backscatter: wireless communication out of thin air,'' in
  \emph{Proc. ACM SIGCOMM}, Aug 2013, pp. 39--50.

\bibitem{Niyato2018surveyambc}
N.~V. Huynh, D.~T. Hoang, X.~Lu, D.~Niyato, P.~Wang, and D.~I. Kim, ``Ambient
  backscatter communications: A contemporary survey,'' \emph{Commun. Surveys
  Tuts.}, Early Access.

\bibitem{Guo2018IoT}
H.~Guo, Q.~Zhang, S.~Xiao, and Y.-C. Liang, ``Exploiting multiple antennas for
  cognitive ambient backscatter communication,'' \emph{IEEE Internet Things
  J.}, vol.~6, no.~1, pp. 765--775, Feb 2019.

\bibitem{Niyato2017RFCRN}
D.~T. Hoang, D.~Niyato, P.~Wang, D.~I. Kim, and Z.~Han, ``Ambient backscatter:
  A new approach to improve network performance for {RF}-powered cognitive
  radio networks,'' \emph{IEEE Trans. Commun.}, vol.~65, no.~9, pp. 3659--3674,
  2017.

\bibitem{GuoAccessSR2019}
H.~Guo, Y.-C. Liang, R.~Long, S.~Xiao, and Q.~Zhang, ``Resource allocation for
  symbiotic radio system with fading channels,'' \emph{IEEE Access}, vol.~7,
  pp. 34\,333--34\,347, 2019.

\bibitem{rzlAccess}
R.~Long, H.~Guo, L.~Zhang, and Y.-C. Liang, ``Full-duplex backscatter
  communications in symbiotic radio systems,'' \emph{IEEE Access}, vol.~7, pp.
  21\,597--21\,608, 2019.

\bibitem{Zhang2018JSAC}
Q.~Zhang, H.~Guo, Y.-C. Liang, and X.~Yuan, ``Constellation learning based
  signal detection for ambient backscatter communication systems,'' \emph{IEEE
  J. Sel. Areas Commun.}, vol.~37, no.~2, pp. 452--463, 2019.

\bibitem{Liaskos2019NeuralNetIRS}
C.~Liaskos, A.~Tsioliaridou, S.~Nie, A.~Pitsillides, S.~Ioannidis, and
  I.~Akyildiz, ``An interpretable neural network for configuring programmable
  wireless environments,'' \emph{arXiv preprint arXiv:1905.02495}, 2019.

\bibitem{Taha2019IRSChannelE}
A.~Taha, M.~Alrabeiah, and A.~Alkhateeb, ``Enabling large intelligent surfaces
  with compressive sensing and deep learning,'' \emph{arXiv preprint
  arXiv:1904.10136}, 2019.

\bibitem{Liaskos2019IRSestimation}
C.~Liaskos, A.~Tsioliaridou, A.~Pitilakis, G.~Pirialakos, O.~Tsilipakos,
  A.~Tasolamprou, N.~Kantartzis, S.~Ioannidis, M.~Kafesaki, and A.~Pitsillides,
  ``Joint compressed sensing and manipulation of wireless emissions with
  intelligent surfaces,'' \emph{arXiv preprint arXiv:1904.10670}, 2019.

\bibitem{Edfors2018IRSMIMOactive}
S.~Hu, F.~Rusek, and O.~Edfors, ``Beyond massive {MIMO}: The potential of data
  transmission with large intelligent surfaces,'' \emph{IEEE Trans. Signal
  Process.}, vol.~66, no.~10, pp. 2746--2758, 2018.

\bibitem{Saad2019activeLISUplink}
M.~Jung, W.~Saad, and G.~Kong, ``Performance analysis of large intelligent
  surfaces ({LISs}): Uplink spectral efficiency and pilot training,''
  \emph{arXiv preprint arXiv:1904.00453}, 2019.

\bibitem{Saad2018activeLISrate}
M.~Jung, W.~Saad, Y.~Jang, G.~Kong, and S.~Choi, ``Performance analysis of
  large intelligence surfaces (liss): Asymptotic data rate and channel
  hardening effects,'' \emph{arXiv preprint arXiv:1810.05667}, 2018.

\bibitem{Griffin2009LinkBudget}
J.~D. Griffin and G.~D. Durgin, ``Complete link budgets for backscatter-radio
  and {RFID} systems,'' \emph{IEEE Antennas Propag. Mag.}, vol.~51, no.~2, pp.
  11--25, 2009.

\bibitem{LiuFu2019Metasurface}
F.~Liu, O.~Tsilipakos, A.~Pitilakis, A.~C. Tasolamprou, M.~S. Mirmoosa, N.~V.
  Kantartzis, D.-H. Kwon, M.~Kafesaki, C.~M. Soukoulis, and S.~A. Tretyakov,
  ``Intelligent metasurfaces with continuously tunable local surface impedance
  for multiple reconfigurable functions,'' \emph{Physical Review Applied},
  vol.~11, no.~4, p. 044024, 2019.

\bibitem{Jochen2007biconvex}
J.~Gorski, F.~Pfeuffer, and K.~Klamroth, ``Biconvex sets and optimization with
  biconvex functions: a survey and extensions,'' \emph{Math. Methods Oper.
  Res.}, vol.~66, no.~3, pp. 373--407, 2007.

\bibitem{LectConvexOpt2001BEN}
A.~Ben-Tal and A.~Nemirovski, \emph{Lectures on modern convex optimization:
  analysis, algorithms, and engineering applications}.\hskip 1em plus 0.5em
  minus 0.4em\relax Siam, 2001, vol.~2.

\bibitem{Boyd2004ConvexOpt}
S.~Boyd and L.~Vandenberghe, \emph{Convex Optimization}.\hskip 1em plus 0.5em
  minus 0.4em\relax Cambridge University Press, 2004.

\bibitem{Wang2019ADMM}
Y.~Wang, W.~Yin, and J.~Zeng, ``Global convergence of {ADMM} in nonconvex
  nonsmooth optimization,'' \emph{Journal of Scientific Computing}, vol.~78,
  no.~1, pp. 29--63, 2019.

\bibitem{Boyd2016Heuristicnonconvexset}
S.~Diamond, R.~Takapoui, and S.~Boyd, ``A general system for heuristic solution
  of convex problems over nonconvex sets,'' \emph{arXiv preprint
  arXiv:1601.07277}, 2016.

\bibitem{Boyd2011ADMM}
S.~Boyd, N.~Parikh, E.~Chu, B.~Peleato, and J.~Eckstein, ``Distributed
  optimization and statistical learning via the alternating direction method of
  multipliers,'' \emph{Foundations and Trends in Machine learning}, vol.~3,
  no.~1, pp. 1--122, 2011.

\bibitem{WMMSE}
Q.~Shi, M.~Razaviyayn, Z.~Luo, and C.~He, ``An iteratively weighted {MMSE}
  approach to distributed sum-utility maximization for a {MIMO} interfering
  broadcast channel,'' \emph{IEEE Trans. Signal Process.}, vol.~59, no.~9, pp.
  4331--4340, 2011.

\bibitem{YuWei2018FP2}
K.~Shen and W.~Yu, ``Fractional programming for communication systems--{Part
  II}: Uplink scheduling via matching,'' \emph{IEEE Trans. Signal Process.},
  vol.~66, no.~10, pp. 2631--2644, 2018.

\bibitem{Kimionis2014enhanceBackefficiency}
J.~Kimionis, A.~Georgiadis, A.~Collado, and M.~M. Tentzeris, ``Enhancement of
  {RF} tag backscatter efficiency with low-power reflection amplifiers,''
  \emph{IEEE Trans. Microw. Theory Techn.}, vol.~62, no.~12, pp. 3562--3571,
  2014.

\end{thebibliography}

\end{document}